\documentclass[11pt]{elsart}
\usepackage{graphicx,amssymb,array,longtable}
\newcommand{\Z}{\mathbb{Z}}
\newcommand{\sig}{\mbox{\boldmath$\sigma$\unboldmath}}
\newcommand{\N}{{\bf N}}
\newcommand{\I}{{\bf I}}
\newcommand{\J}{{\bf J}}
\newcommand{\K}{{\bf K}}
\begin{document}
\begin{frontmatter}
\title{Low temperature expansion of the gonihedric Ising model}
\author{R. Pietig\thanksref{email1}} and 
\author{F.J. Wegner\thanksref{email2}}
\address{Insitut f\"ur Theoretische Physik\\
         Ruprecht-Karls-Universit\"at Heidelberg\\
         Philosophenweg 19, D-69120 Heidelberg, Germany}
\thanks[email1]{email: pietig@pooh.tphys.uni-heidelberg.de}
\thanks[email2]{email: wegner@pooh.tphys.uni-heidelberg.de}
\begin{abstract}
We investigate a model of closed $(d-1)$-dimensional soft-self-avoiding random 
surfaces on a $d$-dimensional cubic lattice. The energy of a surface
configuration is given by $E=J(n_{2}+4k\,n_{4})$, where $n_{2}$ is the number
of edges, where two plaquettes meet at a right angle and $n_{4}$ is the number
of edges, where 4 plaquettes meet. This model can be represented as a 
$\Z_{2}$-spin system with ferromagnetic nearest-neighbour-, antiferromagnetic 
next-nearest-neighbour- and plaquette-interaction. It corresponds to a special
case of a general class of spin systems introduced by Wegner and Savvidy.
Since there is no term proportional to the surface area, the bare surface 
tension of the model vanishes, in contrast to the ordinary Ising model.
By a suitable adaption of Peierls argument, we prove the existence of 
infinitely many ordered low temperature phases for the case $k=0$. A low 
temperature expansion of the free energy in 3 dimensions up to order $x^{38}$ 
($x=\mbox{e}^{-\beta J}$) shows, that for $k>0$ only the ferromagnetic low 
temperature phases remain stable. An analysis of low temperature expansions up 
to order $x^{44}$ for the magnetization, susceptibility and specific heat in 
3 dimensions yields critical exponents, which are in agreement with previous
results.
\end{abstract}
\end{frontmatter}
\section{Introduction}
The so called gonihedric string was introduced by Savvidy et.\ al.\ 
\cite{savvidy1,savvidy2,savvidy3,savvidy4,savvidy5} as a model for closed 
triangulated random surfaces. The action is given as
\begin{eqnarray}
\label{goni_trianguliert}
S & = & \sum_{\mbox{\tiny simple edges}} 
|\pol{X}_{i}-\pol{X}_{j}|\,\theta(\alpha_{ij}) \\ \nonumber
&& +\: k\sum_{\mbox{\tiny common edges}} |\pol{X}_{i}-\pol{X}_{j}|\,
\left(\theta(\alpha_{ij}^{(1)})+\ldots+\theta(\alpha_{ij}^{(r(2r-1))})\right).
\end{eqnarray}
The first term in (\ref{goni_trianguliert}) sums over all edges, where two
triangles meet. The function $\theta(\alpha_{ij})$, which weights the edge
lengths $|\pol{X}_{i}-\pol{X}_{j}|$, depends only on the angle $\alpha_{ij}$
between the neighbouring triangles such that
\begin{itemize}
\item[1.]$\theta(2\pi-\alpha)=\theta(\alpha)$
\item[2.]$\theta(\pi)=0$
\item[3.]$\theta(\alpha)\geq 0.$
\end{itemize} 
If more than two triangles meet at a given edge, contributions from all
pairs of triangles arise in the second term of (\ref{goni_trianguliert}).
Therefore if $k>0$, this term penalizes self-intersections of the surface.
Since $\theta(\pi)=0$, the gonihedric action is subdivision invariant, i.e.\
geometrically nearby surface configurations have similar weights. The action
measures essentially the linear size of the surfaces. Configurations with
long spikes are therefore suppressed. These kind of configurations destroy
the convergence of the partition function for triangulated random surfaces with
area action \cite{ambjorn1}. However, for $k=0$ and 
$\theta(\alpha)=\frac{1}{2}|\pi-\alpha|$, the gonihedric action was shown
to suffer from a similar disease \cite{durhuus}. In this case flat 
``pancake like'' configurations dominate such that the grandcanonical partition 
function fails to converge. Nevertheless numerical simulations of the canonical
ensemble \cite{johnston1,johnston2,johnston3} show flat surfaces. \par
\begin{figure}[htb]
\begin{center}
\begin{tabular}{ccc}
\scalebox{0.82}[1]{\includegraphics[angle=270,width=5cm]{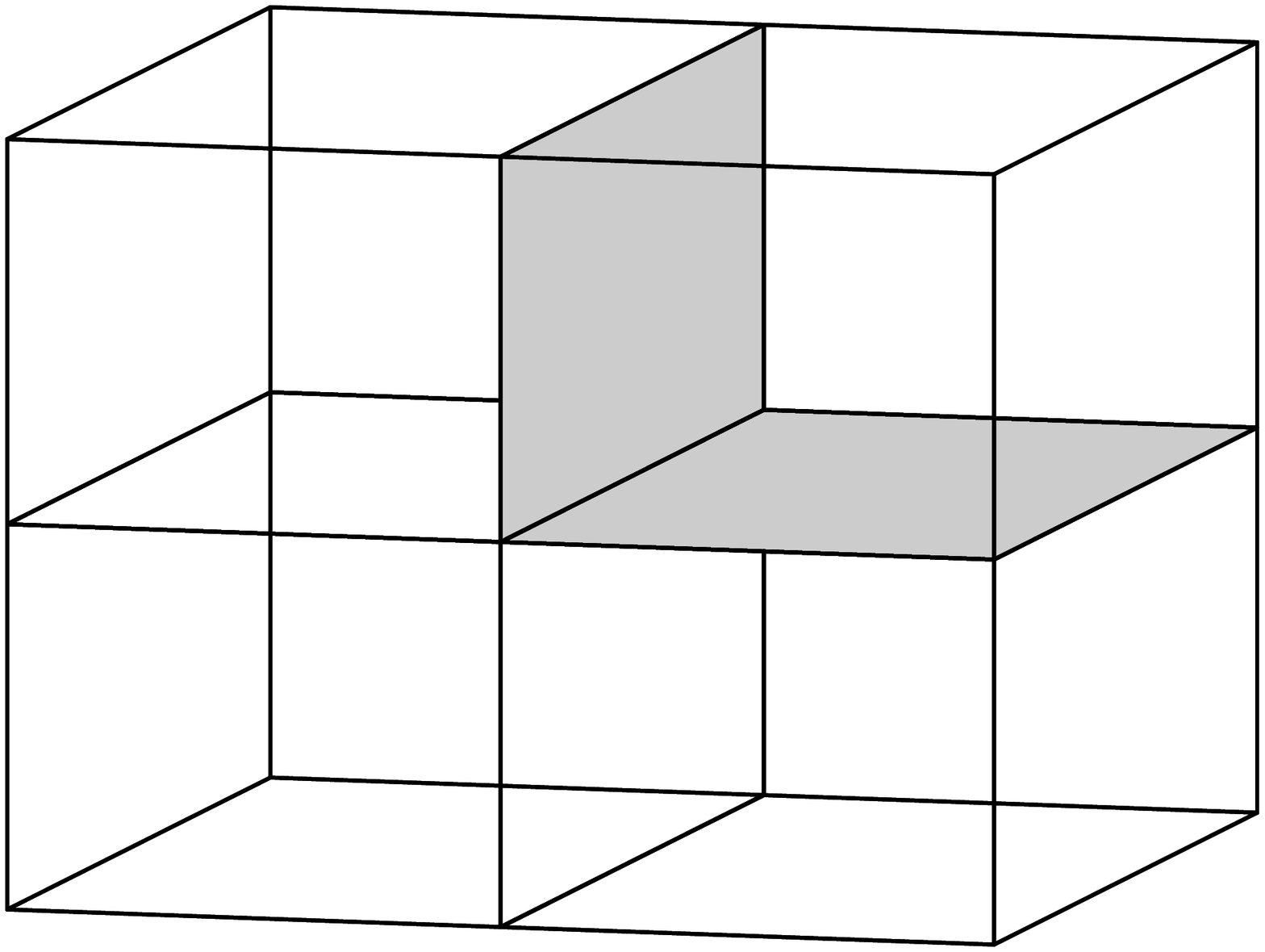}} &
\scalebox{0.82}[1]{\includegraphics[angle=270,width=5cm]{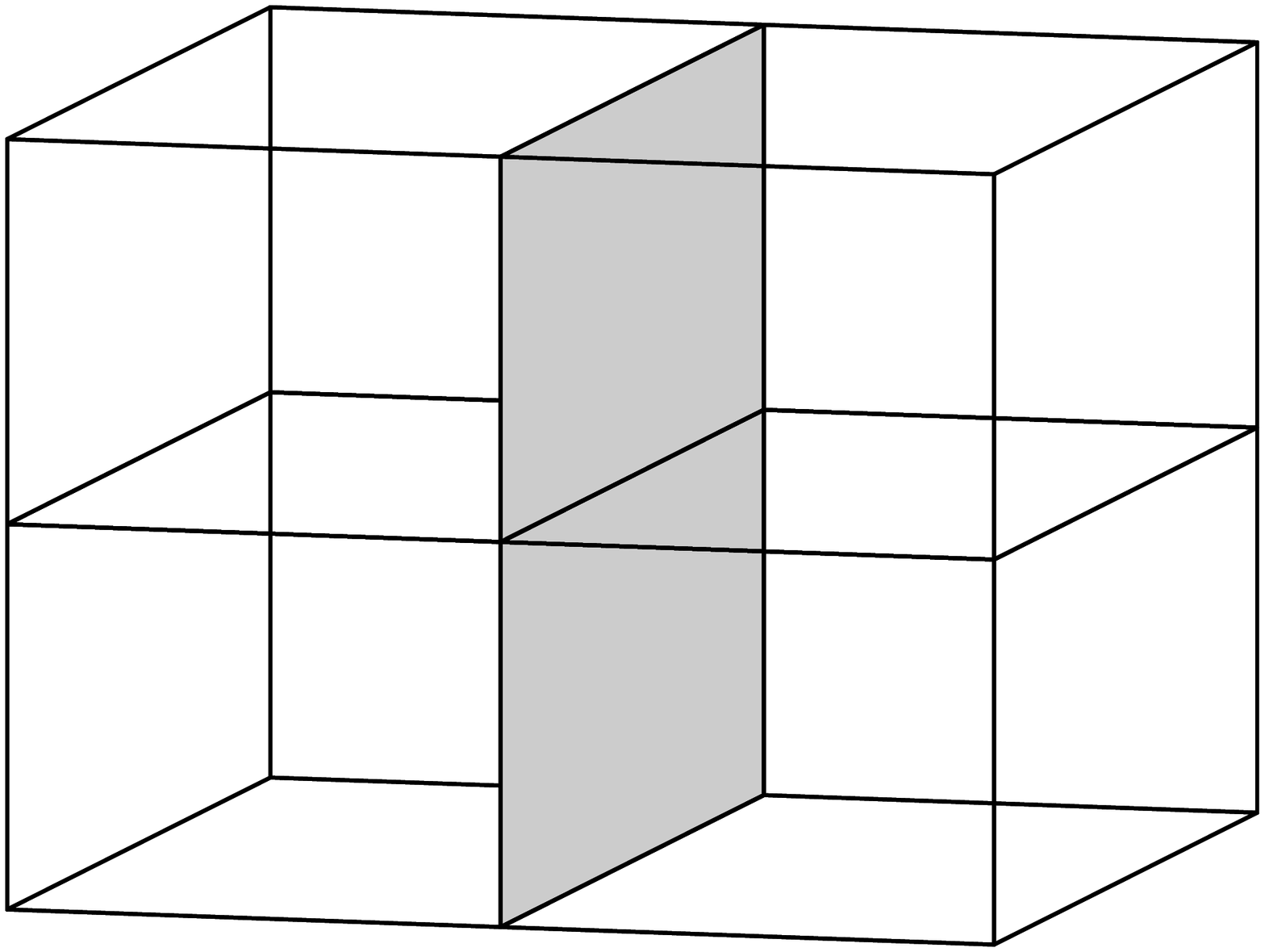}} &
\scalebox{0.82}[1]{\includegraphics[angle=270,width=5cm]{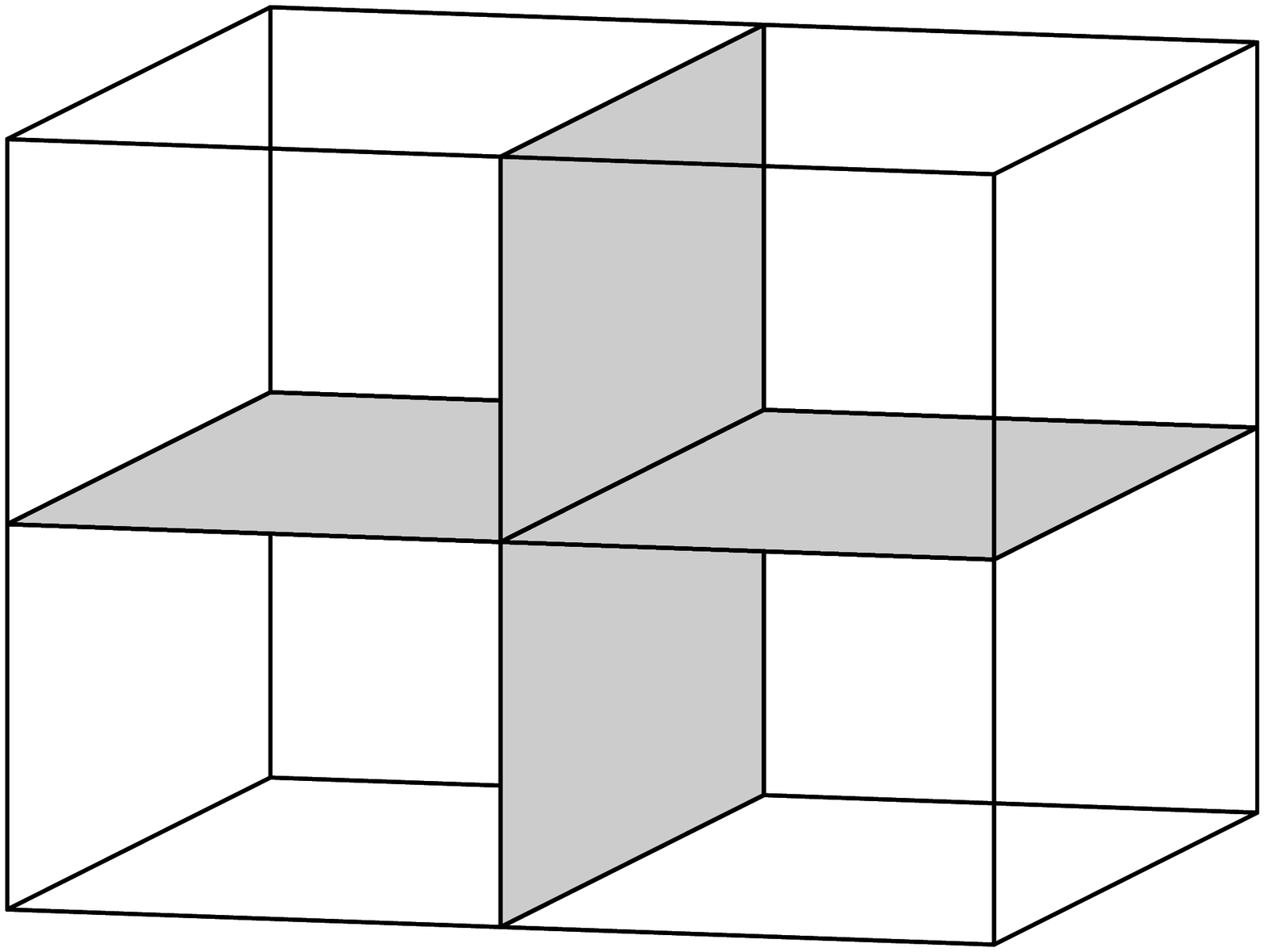}} \\
$E_{\mbox{\tiny edge}}=J$ &
$E_{\mbox{\tiny edge}}=0$ &
$E_{\mbox{\tiny edge}}=4kJ$ 
\end{tabular}
\caption{\label{edges} The total energy can be written as a sum over edge
contributions. ``Simple edges'', where two plaquettes meet perpendicular cost
energy $J$. ``Flat edges'', where two plaquettes of the same orientation
meet, do not cost energy. If four plaquettes meet at a given edge (``double
edge''), the corresponding energy contribution is $4kJ$, since according to
the gonihedric string action four right angles are counted.}
\end{center}
\end{figure}

Another way to define discretized random surface theories is to
consider plaquette surfaces on a cubic lattice. In this approach, not just the
surface, but also the embedding space is discretized. The gonihedric string
can be formulated as a model for plaquette surfaces on a euclidean lattice as 
was shown by Wegner and Savvidy 
\cite{wegner1,wegner2,savvidy6,savvidy7,savvidy8,savvidy9}. If self-overlapping
of the surface is excluded (i.e.\ each plaquette occurs either once or not at 
all in the surface), the model for closed $(d-1)$-dimensional plaquette 
surfaces can be written as a $\Z_{2}$ Ising model. The spins 
$\sigma=\pm 1$ sit on a $d$-dimensional hypercubic lattice and the surface 
consists of all plaquettes of the dual lattice, which separates spins of
opposite sign. The energy of a surface configuration on the lattice is now 
given as 
\begin{equation}
E=\sum_{\mbox{\tiny edges}}E_{\mbox{\tiny edge}}=J(n_{2}+4k\,n_{4}),
\end{equation}
where $n_{2}$ is the number of $(d-2)$-dimensional edges, where two plaquettes 
meet perpendicular and $n_{4}$ is the number of edges, where four plaquettes 
meet (see figure \ref{edges}).
Expressed in terms of spin variables, the Hamiltonian reads
\begin{equation}
\label{hamiltonian}
H=-Jk(d-1)\sum_{<ij>}\sigma_{i}\sigma_{j}
+J\,\frac{k}{2}\sum_{<<ij>>}\sigma_{i}\sigma_{j}
-J\,\frac{1-k}{2}\sum_{[ijkl]}\sigma_{i}\sigma_{j}\sigma_{k}\sigma_{l}.
\end{equation}
It contains ferromagnetic nearest neighbour ($<ij>$), antiferromagnetic
next nearest neighbour ($<<ij>>$) and plaquette terms ($[ijkl]$), where the
corresponding couplings are tuned in a special way. Similar surface models,
where the energy contains an additional term proportional to the surface area,
have been considered earlier 
\cite{karowski1,karowski2,gonnella1,gonnella2,cappi}, in particular in 
connection with amphiphilic systems. However, the special choice of couplings 
as given in (\ref{hamiltonian}), has not been studied explicitly in this 
context. It corresponds to the disorder line as calculated in mean field 
approximation \cite{gonnella2}. In two dimensions, the model can be mapped to 
the eight-vertex model \cite{baxter}. The corresponding weights are however 
different from the exact solvable case. Numerical simulations indicate, that 
the system defined in (\ref{hamiltonian}) does not undergo a phase transition 
for $d=2$ but is in a disordered phase for all $T>0$ \cite{binder,savvidy10}. 
In three dimensions the gonihedric Ising model (\ref{hamiltonian}) has been
studied by means of mean field \cite{johnston4}, Monte Carlo 
\cite{savvidy10,johnston4,johnston5,johnston6} and cluster variation-Pad\'e 
methods \cite{pelizzola1,pelizzola2}. At $k=0$ the model seems to undergo a
first order phase transition \cite{johnston5}, whereas for large $k$ the
transition becomes second order \cite{johnston4,johnston6,pelizzola1}.
Since flat edges, i.e.\ edges where two plaquettes of the same orientation
meet, do not cost energy, the ground state is
highly degenerate. For $k>0$ it consists of all configurations, where only
flat domain walls are present, as long as they do not intersect. The
ground state degeneracy of a finite system of size $L^d$ therefore increases
like $d\,2^L$. If $k=0$, the number of ground states is even higher, since
ground state planes are now allowed to intersect without an additional
energy cost. Thus the degeneracy increases like $2^{dL}$ in this case. 
Moreover, the flipping of a whole $(d-1)$-dimensional spin layer does not change
the energy of any spin configuration if $k=0$ as can easily be seen from
(\ref{hamiltonian}), i.e.\  apart from the global spin-flip symmetry the system
possesses an additional layer-flip symmetry. \par
The outline of the paper is as follows: In section 2 we show by a suitable
extension of Peierls contour method, that the layer-flip symmetry for $k=0$ is 
spontaneously broken at low temperature. For each ground state we find an 
ordered low temperature phase. In section 3 we perform a low temperature 
expansion around all possible ground states of the gonihedric model. The result
indicates, that for $k>0$ only the ferromagnetically ordered low temperature 
phases remain stable, i.e.\ layered phases are thermodynamically suppressed 
at low but non zero temperature. We use Pad\'e approximations to calculate 
critical exponents of the magnetization, susceptibility and specific heat from 
the corresponding low temperature expansions in section 4. Finally we discuss
the surface tension. There is some evidence, that a roughening transition 
occurs below the bulk critical temperature $T_{c}$.
\section{Ordered low temperature phases for \boldmath $k=0$ \unboldmath}
\begin{figure}[ht]
\begin{center}
\includegraphics[angle=270,width=8cm]{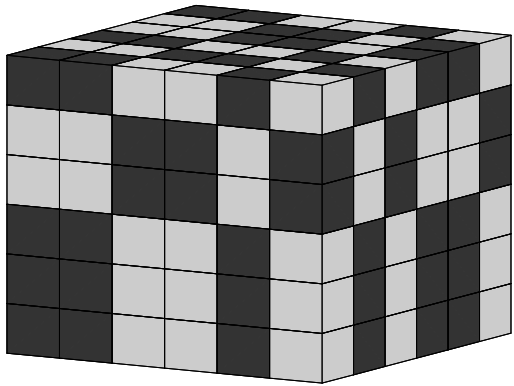}
\caption{\label{rand.ps} Starting from a ferromagnetic ground state, 
further ground states of the gonihedric model for $k=0$ can be constructed by 
swapping $(d-1)$-dimensional spin layers. The figure above visualizes such
a ground state. In order to carry out the Peierls-argument, the boundary spins 
are fixed correspondingly.}
\end{center}
\end{figure}

If $k$ vanishes, then only edges, where two plaquettes meet at a right angle 
cost energy.
These edges are surrounded by an odd number of negative spins. The flipping
of a whole $(d-1)$-dimensional spin layer therefore does not change the energy,
i.e.\ apart from the global $\Z_{2}$ symmetry, the Hamiltonian 
(\ref{hamiltonian}) shows an additional layer-flip symmetry at $k=0$. 
The ground states are the ferromagnetically ordered states and all 
configuration, which are connected with them by layer-flip operations. In the 
following we will show, that for each of these ground states there exists an 
ordered low temperature phase. These phases can be characterized by a non 
vanishing ``spontaneous magnetization'', which for a finite system with $N$ 
spins is defined by
\begin{equation}
\label{mag}
\hat{M}_{N}=\frac{\big<\hat{N}_{+}\big>-\big<\hat{N}_{-}\big>}{N}.
\end{equation}
Here ``$\:\hat{}\:$'' refers to a boundary condition, which singles out one
ground state. $\hat{N}_{-}$ denotes the number of spins, which are 
flipped compared to this ground state and accordingly $\hat{N}_{+}$ the number
of spins, which are unchanged. To show, that $\hat{M}_{N}$ does not vanish
at sufficiently low temperatures, we will use a modified Peierls argument
\cite{peierls,griffiths}. The idea is the following: Consider a finite
system, where the spins at the boundary are fixed, such that one ground state
is favoured (see figure \ref{rand.ps}).
If we now swap a little ``droplet'' of spins inside the volume, the amount
of energy we need will be essentially proportional to the number $l$ of simple 
edges, which were established by swapping the spins. The simple edges form 
connected edge diagrams. Each overturned spin is surrounded with at least one 
such diagram. We can therefore estimate
\begin{equation}
\label{nminus1}
\big<\hat{N}_{-}\big>\leq \sum_{l}
\left(\frac{l}{2d(d-1)}\right)^{\frac{d}{d-2}}p(l), \qquad d\geq 3,
\end{equation}
where $\left(\frac{l}{2d(d-1)}\right)^{d/(d-2)}$ is the
maximum number of spins an edge diagram with $l$ simple edges encloses and
$p(l)$ denotes the probability of occurrence of such a diagram. $p(l)$ can
be further estimated by
\begin{equation}
p(l)\leq g(l)\,\mbox{e}^{-\beta Jl},
\end{equation}
where $g(l)$ is the entropy factor, i.e.\ the number of connected edge diagrams
with $l$ edges. We will show, that $g(l)$ does not grow faster than 
exponentially. Thus (\ref{nminus1}) will be arbitrarily small for sufficiently
large $\beta$. The same type of argument was used in \cite{pietig} to show
the existence of a phase transition in the gonihedric model for $k>0$. The
argument given there however contains a flaw, since the edge diagrams are
not independent from each other, if $k>0$\footnote{For details see \cite{diss}}.
We will now proceed with the details for the case $k=0$. \par
\begin{figure}[htb]
{\renewcommand{\arraystretch}{0.5}
\begin{tabular}{cccc} &&&\\
2-Vertex & 3-Vertex & 4-Vertex & 5-Vertex \\
\scalebox{0.82}[1]{\includegraphics[angle=270,width=3.5cm]{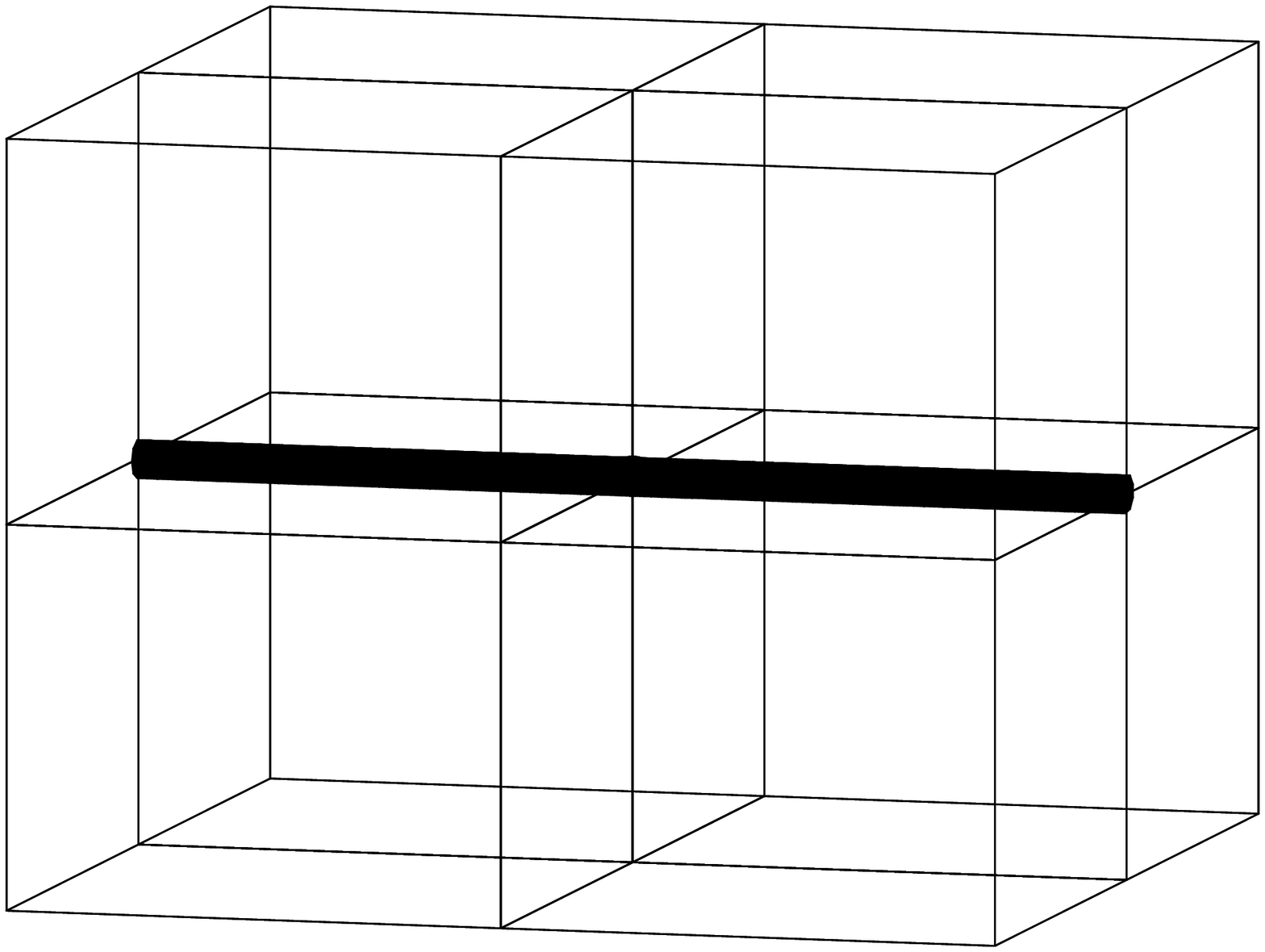}} &
\scalebox{0.82}[1]{\includegraphics[angle=270,width=3.5cm]{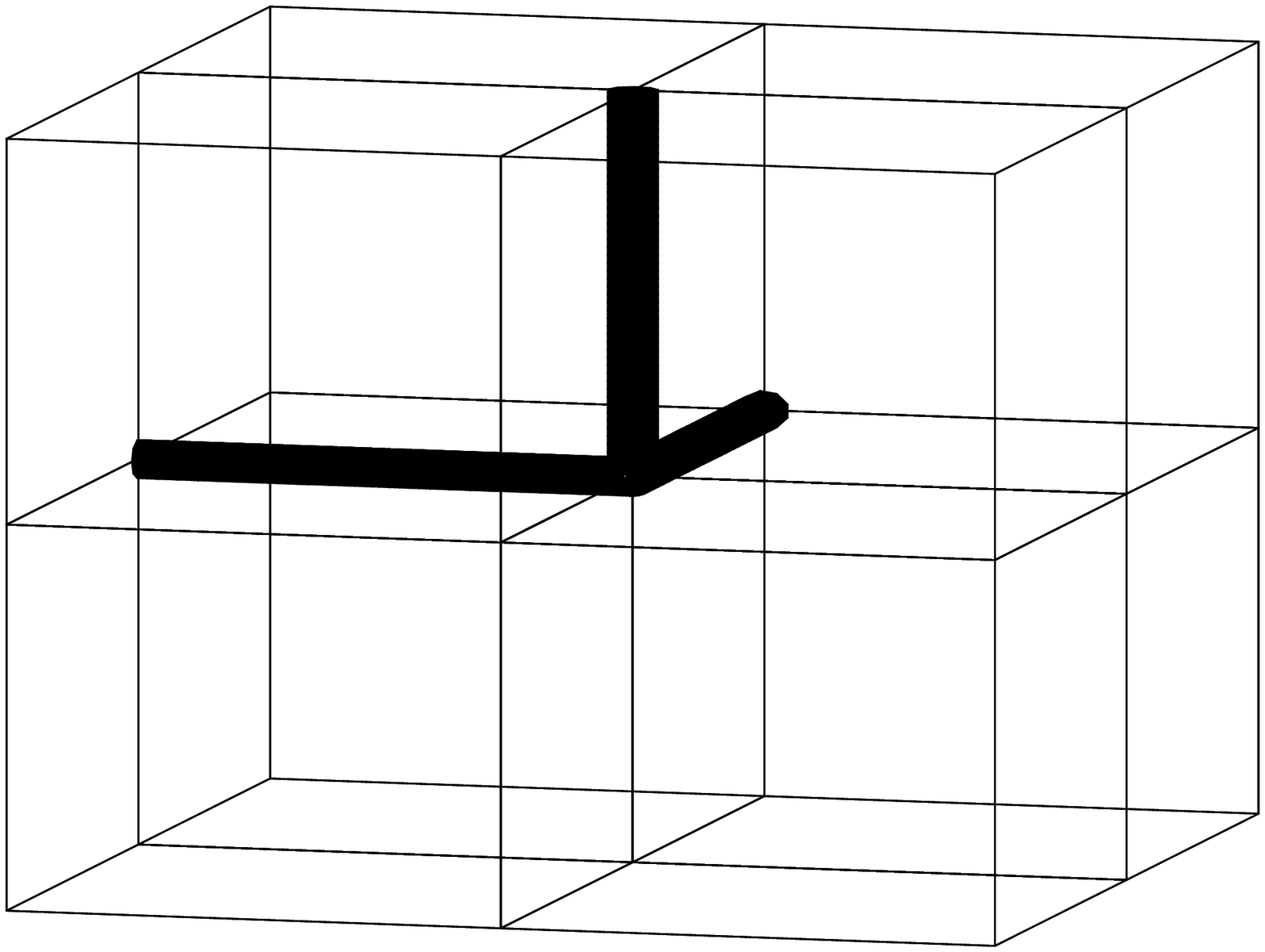}} &
\scalebox{0.82}[1]{\includegraphics[angle=270,width=3.5cm]{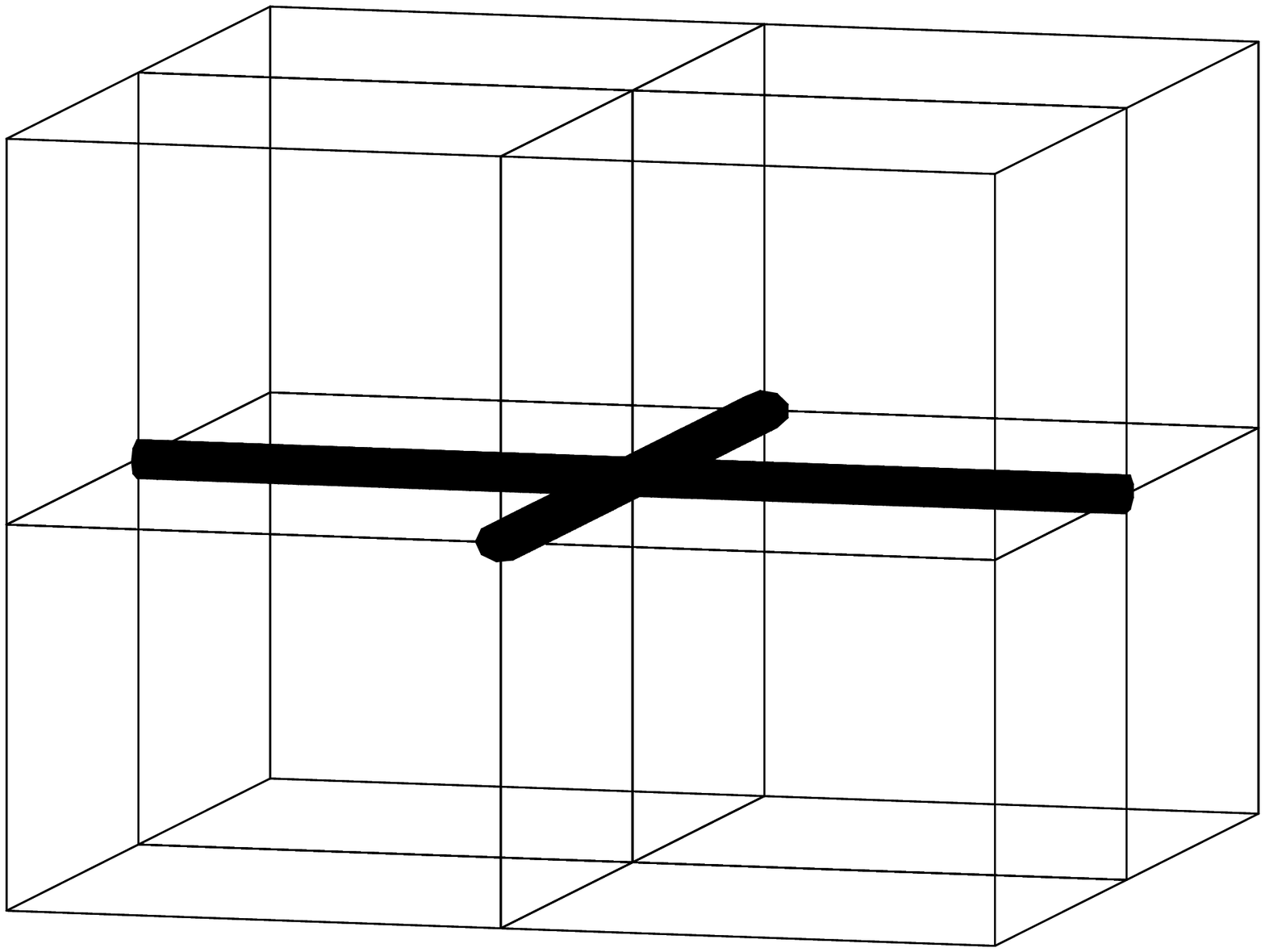}} &
\scalebox{0.82}[1]{\includegraphics[angle=270,width=3.5cm]{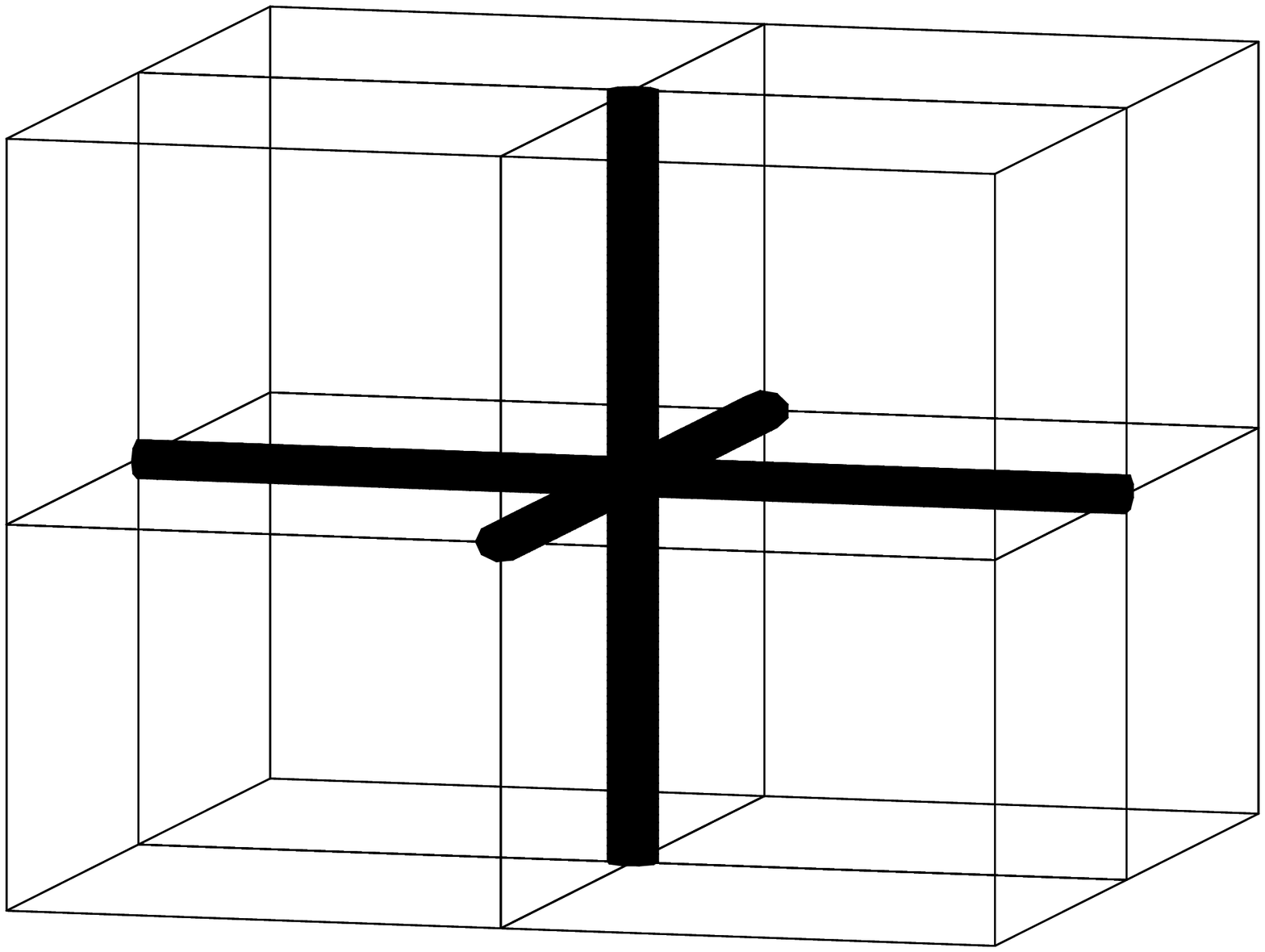}} 
\end{tabular}}
\caption{\label{vertex} The possible edge configurations around a vertex
are determined by the eight surrounding spins. Only four types of
configurations can arise (modulo rotations) as shown above.}
\end{figure}
\begin{figure}[hbt]
{\renewcommand{\arraystretch}{1}
\setlength{\tabcolsep}{1mm}
\begin{tabular}{|c|c||c|c||c|c||c|c||c|c|} \hline
\multicolumn{2}{|c||}{
\scalebox{0.82}[1]{\includegraphics[angle=270,width=3cm]{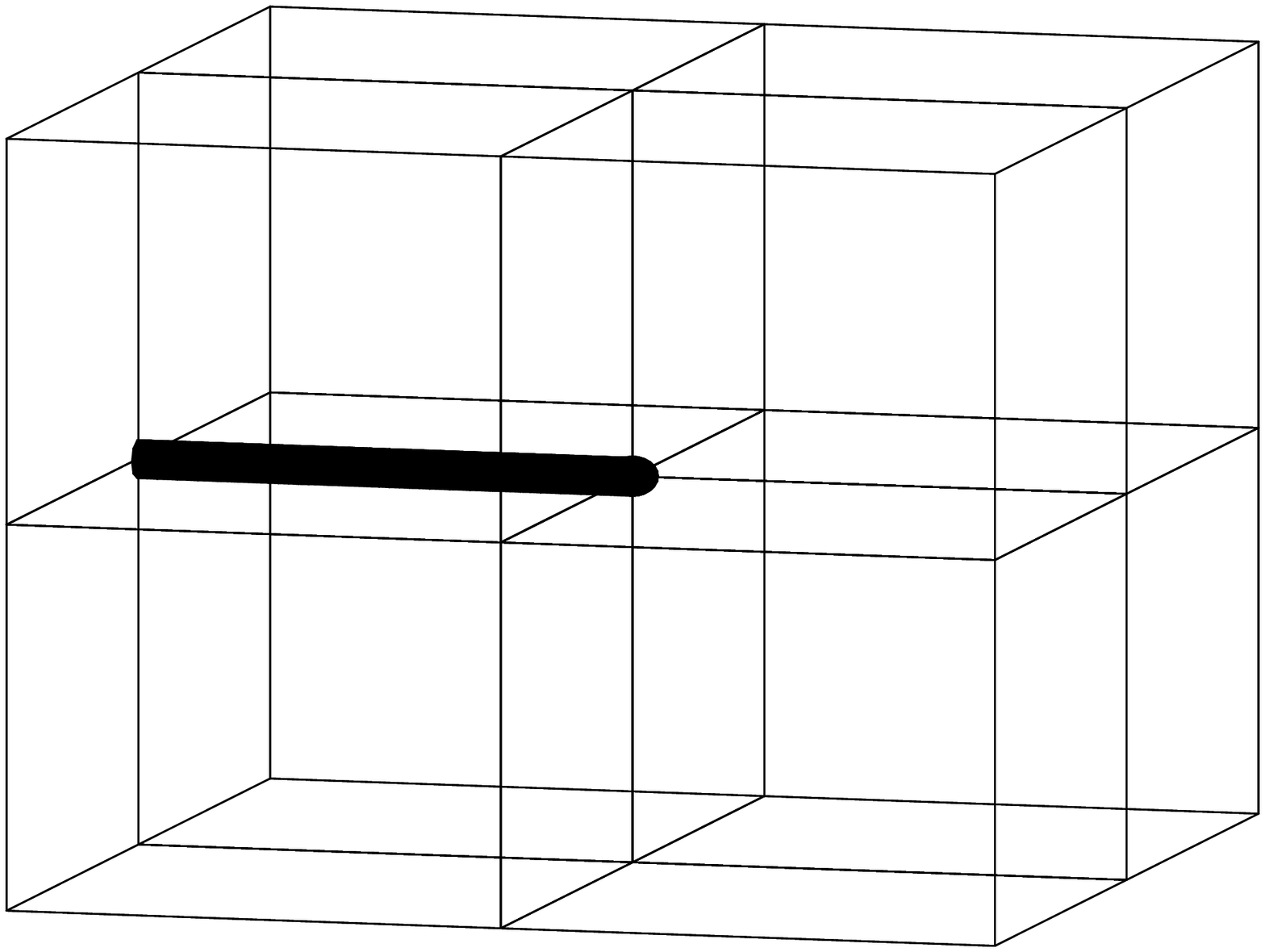}}} &
\multicolumn{2}{|c||}{
\scalebox{0.82}[1]{\includegraphics[angle=270,width=3cm]{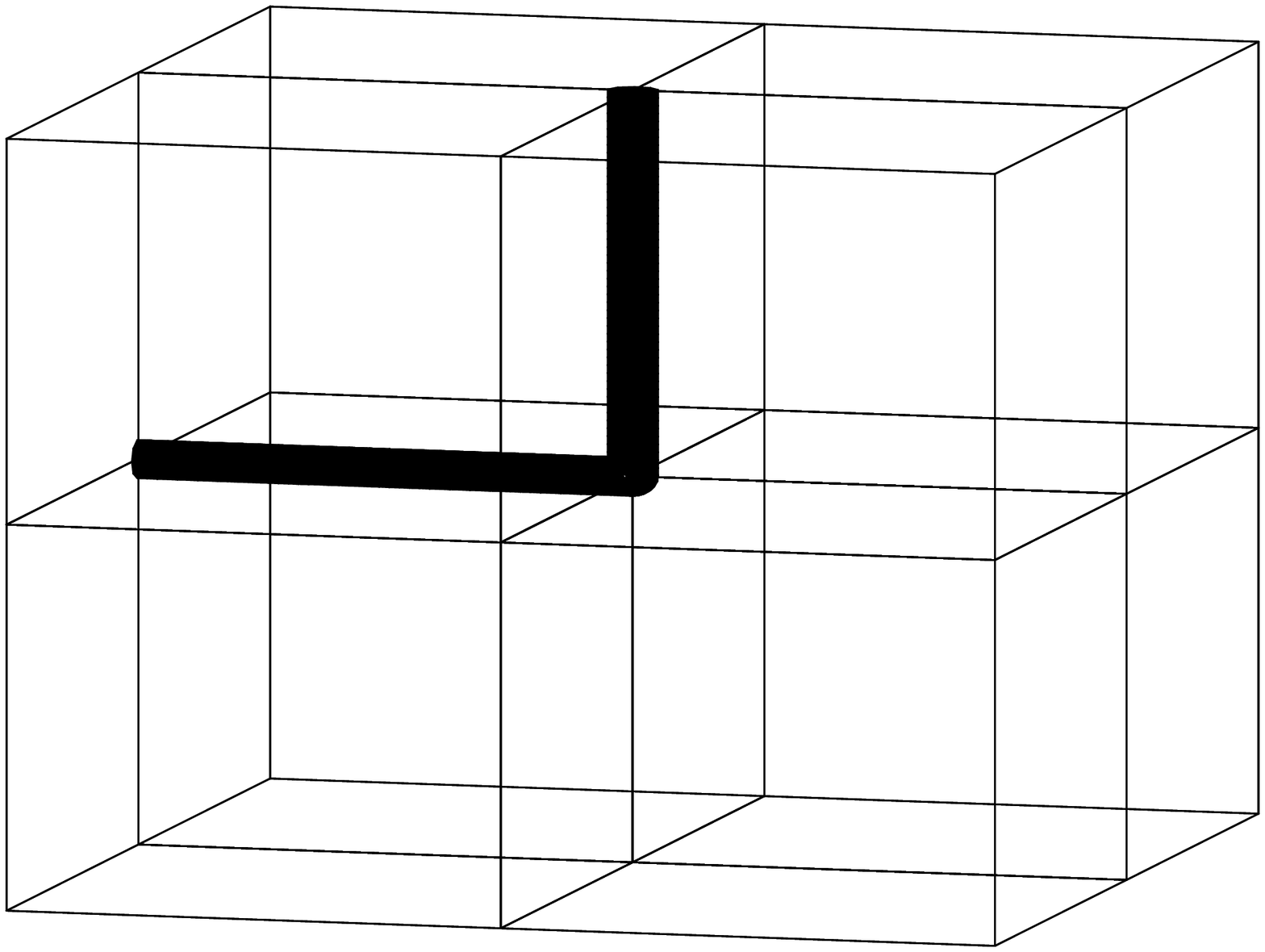}}} &
\multicolumn{2}{|c||}{
\scalebox{0.82}[1]{\includegraphics[angle=270,width=3cm]{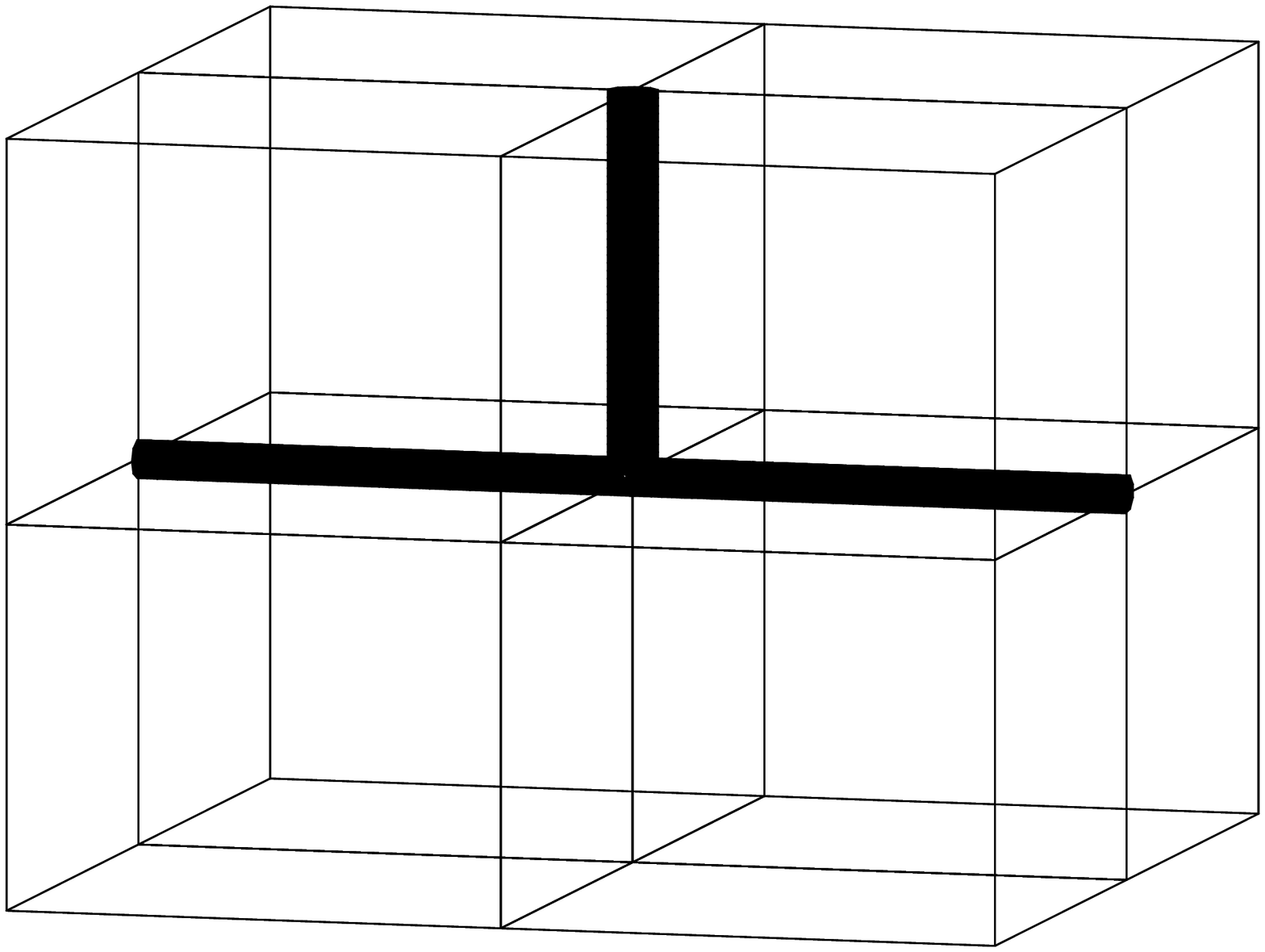}}} &
\multicolumn{2}{|c||}{
\scalebox{0.82}[1]{\includegraphics[angle=270,width=3cm]{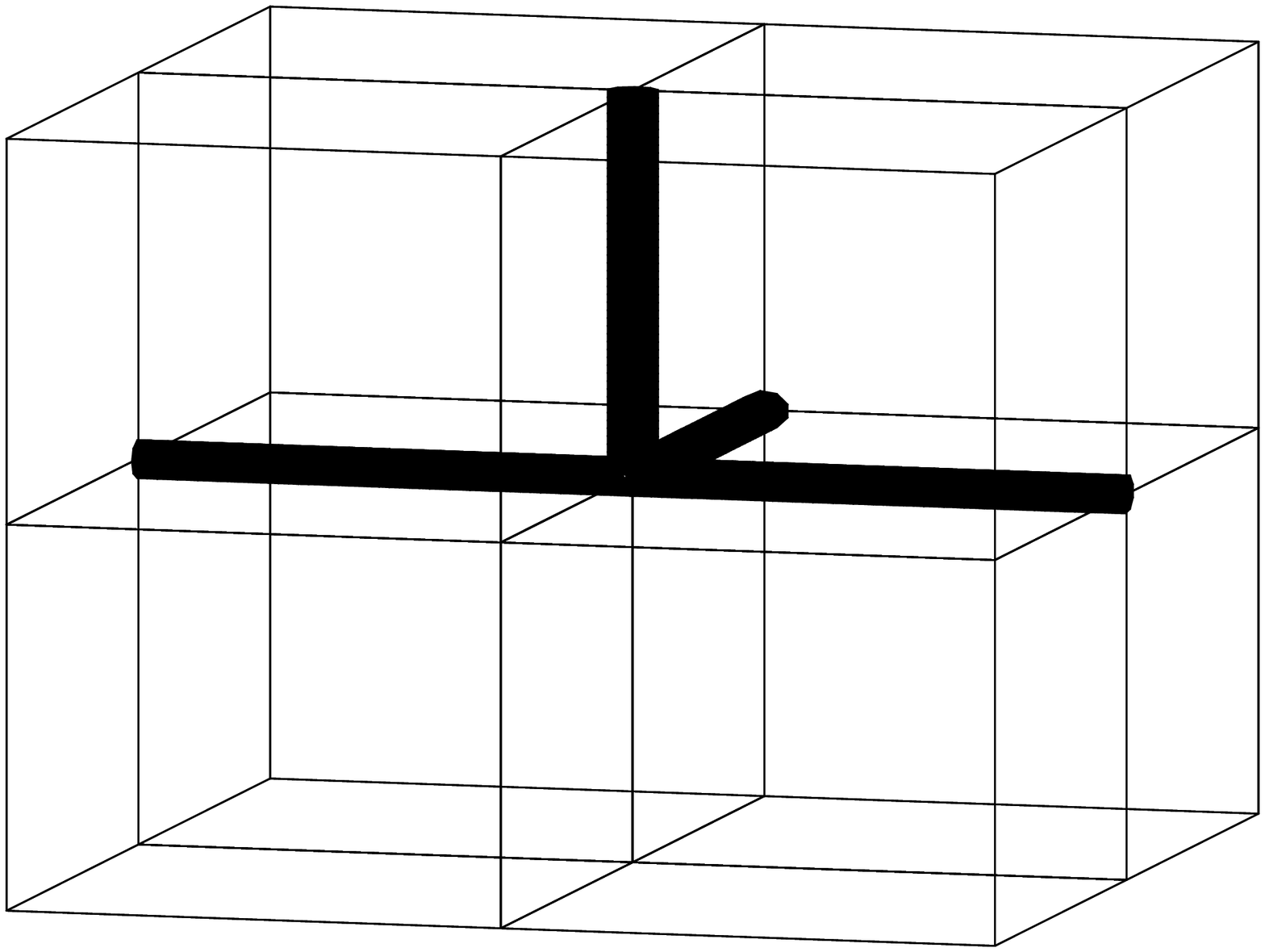}}} &
\multicolumn{2}{|c|}{
\scalebox{0.82}[1]{\includegraphics[angle=270,width=3cm]{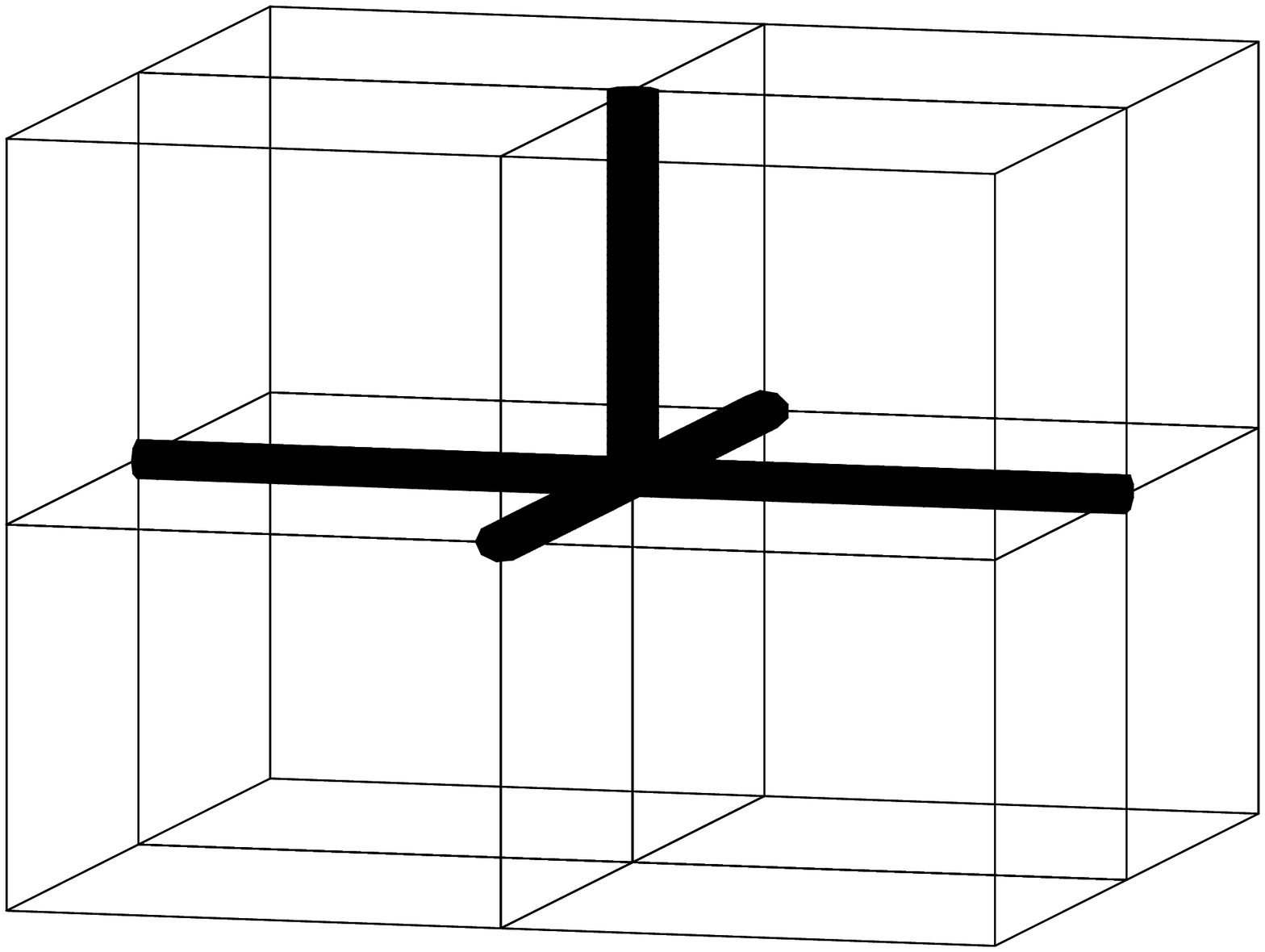}}} 
\\ \hline \hline
$k$ & $n(k)$ & $k$ & $n(k)$ & $k$ & $n(k)$ & $k$ & $n(k)$ & $k$ & $n(k)$
\\ \hline \hline
1 & 1 & 1 & 2 & 1 & 1 & 1 & 0 & 1 & 1 \\ \hline
2 & 4 & 2 & 1 & 2 & 0 & 2 & 1 & 2 & 0 \\ \hline
3 & 2 & 3 & 0 & 3 & 1 & 3 & 0 & 3 & 0 \\ \hline
4 & 0 & 4 & 1 & 4 & 0 & 4 & 0 & 4 & 0 \\ \hline
5 & 1 & 5 & 0 & 5 & 0 & 5 & 0 & 5 & 0 \\ \hline
\end{tabular}}
\caption{\label{choice} The figure shows all possibilities of incomplete
vertices (modulo rotations). $n(k)$ is the number of choices to complete
the vertex by adding $k$ edges.} 
\end{figure}

Consider a finite system with $N$ spins in its interior and fix the spins
at the boundary, so that they belong to a ground state, as indicated in figure
\ref{rand.ps}. For a configuration $\sig$ of the spins inside the volume let
$C[\sig]$ be the set of all ($(d-2)$-dimensional) edges of the dual lattice, 
where two ($(d-1)$-dimensional) plaquettes of the domain wall meet at a right 
angle. Each of these edges is surrounded by four spin, whose product is -1.
Two edges in $C[\sig]$ are connected, if they have at least one 
($(d-3)$-dimensional) vertex in common. $C[\sig]$ can be uniquely decomposed
into connected parts:
\begin{equation}
C[\sig]=C_{1}^{l_{1}}\cup\ldots\cup C_{n}^{l_{n}},
\end{equation}  
where $C_{i}^{l}$ denotes a connected component with $l_{i}$ edges. The total
energy $E[\sig]$ can be written as
\begin{equation}
E[\sig]=\sum_{i=1}^{n}J\,l_{i}.
\end{equation}
Let $g(l)$ be the number of connected edge diagrams with $l$ edges. For each
diagram we introduce a variable $\chi_{i}^{l}, i=1,\ldots,g(l)$,
such that 
\begin{equation}
\chi_{i}^{l}[\sig]=\left\{
\begin{array}{rcl}
1 && \parbox[t]{4.5cm}{if the corresponding edge diagram occurs in $\sig$} 
                     \\[5mm]
0 && \mbox{otherwise}.
\end{array}\right.
\end{equation}
Since each spin, which is flipped compared to the ground state, is surrounded
with at least one connected edge diagram, we can estimate
\begin{equation}
\hat{N}_{-}[\sig]\leq 
\sum_{l}\;\left(\frac{l}{2d(d-1)}\right)^{\frac{d}{d-2}}\:
\sum_{i=1}^{g(l)}\chi_{i}^{l}[\sig].
\end{equation}
Thus for the expectation value of the number of the overturned spins we obtain:
\begin{equation}
\label{hNminus1}
\big<\hat{N}_{-}\big>_{N}\: \leq 
\sum_{l}\;\left(\frac{l}{2d(d-1)}\right)^{\frac{d}{d-2}}\:
\sum_{i=1}^{g(l)} \big<\chi_{i}^{l}\big>_{N}.
\end{equation}  
If we number the spin configurations, which contain $C_{i}^{l}$ by
$\sig_{1},\ldots,\sig_{k}$, the expectation value $\big<\chi_{i}^{l}\big>_{N}$
can be written as:
\begin{eqnarray}
\label{hchi}
\big<\chi_{i}^{l}\big>_{N} & = & \frac{1}{Z_{N}}\sum_{j=1}^{k} 
\mbox{e}^{-\beta E[\mbox{\boldmath$\scriptstyle\sigma_{j}$\unboldmath}]},\\[3mm]
\mbox{where \hspace{1cm}} && \nonumber \\
Z_{N} & = & \sum_{\mbox{\boldmath$\scriptstyle\sigma$\unboldmath}}
\mbox{e}^{-\beta E[\mbox{\boldmath$\scriptstyle\sigma$\unboldmath}]}.
\end{eqnarray}
For each $\sig_{j}$ there exists a unique configuration $\sig_{j}^{*}$, which
does not contain $C_{i}^{l}$ but leaves all other edge diagram unchanged, i.e.\ 
\begin{equation}
E[\sig_{j}^{*}]=E[\sig_{j}]-J\,l.
\end{equation}
The configurations $\sig_{1}^{*},\ldots,\sig_{k}^{*}$ are pairwise
different. By restricting the partition function to these configurations,
we can estimate:
\begin{equation}
Z_{N} \geq \sum_{j=1}^{k} 
\mbox{e}^{-\beta E[\mbox{\boldmath$\scriptstyle\sigma_{j}^{*}$\unboldmath}]} = 
\mbox{e}^{\beta Jl} \,\sum_{j=1}^{k} 
\mbox{e}^{-\beta E[\mbox{\boldmath$\scriptstyle\sigma_{j}$\unboldmath}]}.
\end{equation}
Together with (\ref{hchi}) this yields
\begin{equation}
\big<\chi_{i}^{l}\big>_{N} \:\:\leq \mbox{e}^{-\beta Jl}.
\end{equation}
Hence the expectation value of the number of overturned spins is bounded by
\begin{equation}
\label{hNminus2}
\big<\hat{N}_{-}\big>_{N}\: 
\leq \sum_{l}\left(\frac{l}{2d(d-1)}\right)^{\frac{d}{d-2}}
g(l) \;\mbox{e}^{-\beta Jl}.
\end{equation}
To proceed with the argument, we need an upper bound for $g(l)$, the number
of connected edge diagrams with $l$ edges. The edges in those diagrams are
connected via ($(d-3)$-dimensional) vertices. What kind of edge configurations 
can occur at a vertex is determined by the possible configurations of the eight
surrounding spins. As shown in figure \ref{vertex}, only four types of vertices
can arise.
With the following method all possible connected edge diagrams can be 
constructed:
\begin{itemize}
\item[1.] Number all Vertices ($(d-3)$-dimensional plaquettes) of the lattice.
\item[2.] Choose one of the $N$ lattice points, e.g.\ 
$r=(r_{1},\ldots,r_{d}), r_{i}\in \Z$ and attach the edge 
$k=\{(r_{1},r_{2},\lambda_{3},\ldots,\lambda_{d}) \:|\:
r_{i}\leq\lambda_{i}\leq r_{i}+1\}$. This is possible, since edges of all
orientations occur in every possible diagram.
\item[3.] Imagine that a connected subdiagram exists already. Consider all
vertices in that diagram, where the configuration of the surrounding edges
is not allowed (see figure \ref{choice}). Choose the one with the lowest 
number and attach further edges, such that an allowed configuration results.
\item[4.] Repeat constructions step 3, until all $l$ edges are attached.
\end{itemize}
The maximum number of outcomes of the procedure given above is an upper bound
for $g(l)$. For construction step 2 there are $N$ possibilities. Each time we
perform constructions step 3, at least one edge is added. The number of choices
to complete the vertex depends on the edge configuration already present. In 
figure \ref{choice}, we list all possibilities which can arise together with 
the number of choices $n(k)$ to attach $k$ edges.
Let $n_{\mbox{\tiny max}}(k)$ be the maximum number of choices to attach $k$
edges. The function
\begin{equation}
f_{r}(x):=N\,x\left(\;\sum_{k=1}^{5}n_{\mbox{\tiny max}}(k)\,x^{k}\right)^{r},
\end{equation}
can now be interpreted as a generating function, which counts all connected
edge diagrams with $l$ edges at least once, which can be constructed by
repeating constructions step 3 r-times, i.e. the coefficient of $x^{l}$ in the 
series expansion of $f_{r}(x)$ is an upper bound for the number of those 
diagrams. Therefore an upper bound for $g(l)$ is provided by
\begin{equation}
g(l)\leq \frac{1}{l!}\left(\frac{d}{dx}\right)^{l}\left.
\left(\;\sum_{r=0}^{\infty}f_{r}(x)\right)\:\right|_{x=0}.
\end{equation}
The values $n_{\mbox{\tiny max}}(k)$ can be read off from figure \ref{choice}.
We find:
\begin{eqnarray}
\label{gl1}
g(l) & \leq & \frac{1}{l!}\left(\frac{d}{dx}\right)^{l}\left.
\left(\frac{N\,x}{1-(2x+4x^2+2x^3+x^4+x^5)}\right)\:\right|_{x=0} \\
& =: & \frac{1}{l!}\left(\frac{d}{dx}\right)^{l}\left. G(x) \:\right|_{x=0}.
\end{eqnarray}
$G(x)$ can be written as:
\begin{equation}
G(x)=\frac{N\,x}{1-(2x+4x^2+2x^3+x^4+x^5)}=N\,
\sum_{k=1}^{5}\frac{r_{k}}{x-x_{k}}.
\end{equation}
Putting this expression in (\ref{gl1}), we obtain
\begin{equation}
g(l)\leq N\,\sum_{k=1}^{5}\left(-\frac{r_{k}}{x_{k}}\right)
\left(\frac{1}{x_{k}}\right)^{l}.
\end{equation}
The increase of $g(l)$ is dominated by the smallest 
$|x_{k}|=:x_{\mbox{\tiny min}}$.
Hence we can further estimate:
\begin{equation}
g(l)\leq N\,\left(\sum_{k=1}^{5}\left|\frac{r_{k}}{x_{k}}\right|\right)
\left(\frac{1}{|x_{m}|}\right)^{l}\leq
N\,0.67\,(3.39)^{l}.
\end{equation}
Substituting this result in (\ref{hNminus2}), we find an upper bound for the
density of overturned spins:
\begin{equation}
\label{hNminus3}
\frac{\big<\hat{N}_{-}\big>_{N}}{N}\: 
\leq \sum_{l}\left(\frac{l}{2d(d-1)}\right)^{\frac{d}{d-2}}
0.67\,(3.39)^{l} \;\mbox{e}^{-\beta Jl}.
\end{equation}
This inequality remains valid in the thermodynamical limit. The sum on the
right hand side approaches zero, if $\beta$ tends to infinity. Therefore, for
$\beta>\beta_{c}$ it will be smaller than $\frac{1}{2}$, which implies a 
non-zero spontaneous magnetization. This completes the proof. Note that
the estimate above is only valid for $d\geq 3$. Indeed in two dimensions,
the model for $k=0$ is equivalent to the 1-dimensional ordinary Ising model
\cite{wegner2}. Thus no transition occurs in this case. \par
The upper bound for $\big<\hat{N}_{-}\big>_{N}$ given above can be used to
calculate an upper bound for $\beta_{c}$. We find
\begin{eqnarray*}
\beta_{c}J\leq 1.43 &\quad & \mbox{for} \quad d=3, \\
\beta_{c}J\leq 1.33 &\quad & \mbox{for} \quad d=4. 
\end{eqnarray*}
The upper bound for $\beta_{c}J$ in 3 dimensions is consistent with values
obtained from simulations \cite{johnston4,johnston5} 
($\beta_{c}J\approx 0.505$),
cluster variation-Pad\'e approximations \cite{pelizzola1}  
($\beta_{c}J\approx 0.550$) and mean field calculations \cite{johnston4}
($\beta_{c}J\approx 0.325$). \par
If $k>0$ a similar proof can not carried through, since in this case
the connected edge diagrams are not independent from each other, i.e. given
a configuration $\sig$ which contains a certain edge diagram $C_{i}^{l}$,
the corresponding configuration $\sig^{*}$ does not necessarily fulfil the
condition
\begin{equation}
E[\sig^{*}]=E[\sig]-E[C_{i}^{l}],
\end{equation}
where $E[C_{i}^{l}]$ denotes the energy contribution of the connected edge
diagram. In going over from $\sig$ to $\sig^{*}$, the energy can even increase
by an amount $\sim l^2$.
\section{Low temperature expansion of the free energy}
If $k>0$ then the Hamiltonian (\ref{hamiltonian}) is no longer invariant under
layer-flip operations. Only the global $\Z_{2}$ spin-flip symmetry remains.
However flat domain walls still do not cost energy. Thus apart from the
ferromagnetically ordered ground states, there are layered ground states
containing flat domain walls which do not intersect. As was already mentioned
in \cite{pelizzola1}, the ground state contributions $Z^{(g)}$ of the
total partition function
\begin{equation}
Z=\sum_{g}Z^{(g)}
\end{equation}
are not degenerate in contrast to the $k=0$ case. Here $Z^{(g)}$ contains the
ground state $g$ and all configurations, which differ from $g$ by a ``small''
number of spins. The lowest excitation is realized by a configuration, where
one spin is flipped compared to the ground state. Since this configuration
contains 12 simple edges (elementary cube), its energy is $12J$, independent
of the underlying ground state. The next higher excitation is obtained by
swapping an additional nearest neighbour spin. The energy of this configuration
however depends on the ground state: If there is no flat domain wall present
between those two spins, the energy is 16J. Otherwise four additional double
edges contribute, i.e. the energy amounts to $16J+16Jk$. Thus to lowest order,
layered low temperature phases seem to be energetically disfavoured 
\cite{pelizzola1}. In this section we will quantify this effect. For this we
perform a low temperature expansion of the free energy for the three 
dimensional case to estimate the magnitude of each ground state contribution 
$Z^{(g)}$. This analysis indicates, that at low but non-zero temperature the 
occurrence of layered low temperature phases is indeed thermodynamically 
suppressed. \par
\begin{figure}[htb]
\begin{center}
\scalebox{0.8}[1]{\includegraphics[angle=270,width=9cm]{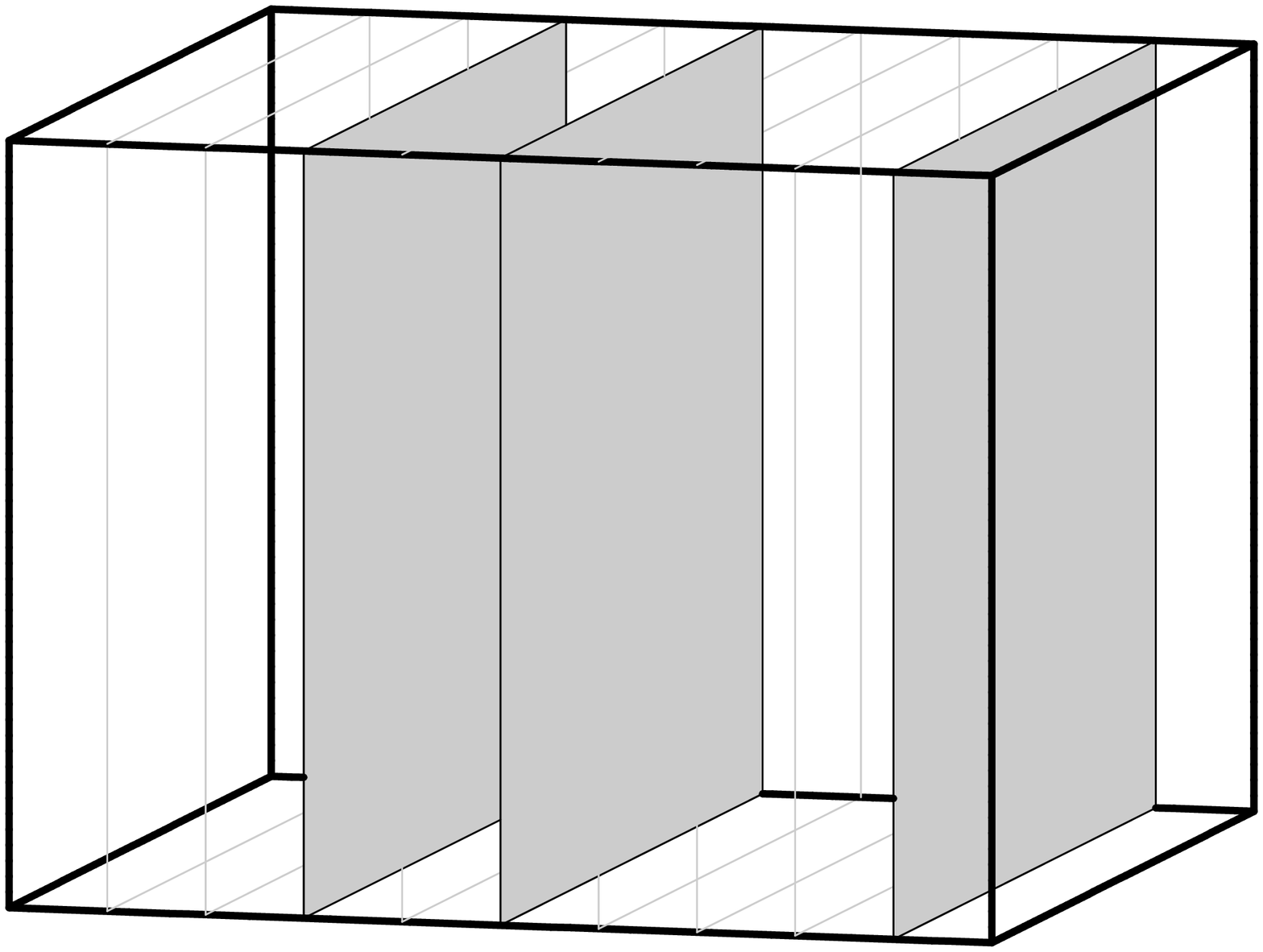}} \\
$L=10$, \qquad $p(g)=(0,0,1,0,1,0,0,0,1)$
\caption{\label{tupel.ps} The picture above displays ground state layers
within a finite volume of linear length $L=10$. The sequence of this layers
in positive direction (from left to right) is described by a $(L-1)$-tupel $p$.}
\end{center}
\end{figure}

Consider a finite system of volume $V=L^{3}$. By $\N$ we denote the set of the 
$N=L^{3}$ spins inside $V$. Each excitation of a given ground state g is 
defined by the subset $\I\subseteq \N$ of spins, which are swapped compared to 
g. The excitation energy $E^{(g)}(\I)$ depends on g. We define the ground state 
contributions $Z^{(g)}_{N}$ by:
\begin{eqnarray}
\label{grundzustandsb}
Z^{(g)}_{N} & := & \sum_{\I\subseteq\N}z^{(g)}(\I), \\[1mm]
\mbox{where \hspace{1cm}} && \nonumber \\[1mm]
\quad z^{(g)}(\I) & := & \mbox{e}^{-\beta E^{(g)}(\I)}.
\end{eqnarray} 
Ground states g which can be mapped onto each other by a rotation or a global
spin flip yield the same $Z^{(g)}_{N}$. We therefore write:
\begin{equation}
\label{zerlegung}
Z_{N}=2\Big(Z_{N}^{(0)}+3\sum_{g=1}^{2^{L-1}-1}Z_{N}^{(g)}\Big),
\end{equation}
where $g$ refers to ground states modulo rotations and global spin-flip and
$g=0$ denotes the ferromagnetically ordered ground state. The logarithm of
$Z^{(g)}_{N}$ can be written as:
\begin{equation}
\label{clustersumme}
\mbox{ln}(Z^{(g)}_{N}) = \sum_{\I\subseteq\N}c^{(g)}(\I),
\end{equation}
where the cluster contributions $c^{(g)}(\I)$ are defined by
\begin{equation}
c^{(g)}(\I) = \sum_{\J\subseteq\I}(-1)^{(|\I|-|\J|)}\:
\mbox{ln}\Big(\sum_{\K\subseteq\J}z^{(g)}(\K)\Big).
\end{equation}
As can easily be shown, these cluster contributions vanish, if the cluster
$\I$ is not ``connected''. Here we call two spins connected, if their
elementary cubes have at least one edge in common. In general two spin clusters
$\I_{1},\I_{2}$, which can be mapped onto each other by a translation will
contribute differently, unless the translation is parallel to the ground state
planes. To take care of this restricted translation invariance we introduce 
structural coefficients, which characterize the ground states such that the
right hand side of (\ref{clustersumme}) can be expanded systematically: 
For each ground state $g$ there exists a unique $(L-1)$-tupel
$p(g)=(p_{1},\ldots,p_{L-1})$, $p_{i}=0,1$, which characterizes the sequence
of ground state planes say in positive direction. $p_{i}=1$ means, that a
ground state plane is present (see figure \ref{tupel.ps}).
For $q=(q_{1},\ldots,q_{n})$, $q_{i}=0,1$, $n<=L-1$ and
\begin{equation}
T:=\{k\in\mathbb{N} \;|\; q_{1}=p_{k},\ldots,q_{n}=p_{k+n-1}\}.
\end{equation}
we define:
\begin{equation}
a_{q}:=\mbox{Number of elements in $T$}.
\end{equation}
Thus $a_{q}(g)$ is the number of translations, such that the sequence $q$ 
matches with a subsequence of $p(g)$. If $q$ does not have any elements, we set 
$a(g)=L$. The structural coefficients are not independent from each other, e.g.\
the following equations hold:
\begin{equation}
\label{constraint}
\begin{array}{rcl}
a(g) & = & L \\
a_{0}(g)+a_{1}(g) & = & L-1 \\
a_{00}(g)+a_{01}(g)+a_{10}(g)+a_{11}(g) & = & L-2 \\
& \vdots &
\end{array}
\end{equation}
Using the structural coefficients, we can now write:
\begin{equation}
\label{darstellung}
\lim_{L\to\infty}\frac{\mbox{ln}(Z_{N}^{(g)})}{N}=
\lim_{L\to\infty}\frac{1}{L}\Big(a(g)f+a_{0}(g)f_{0}+a_{1}(g)f_{1}
+a_{00}(g)f_{00}+\ldots\Big).
\end{equation}
In this expression $f$ denotes the sum of all cluster contributions of widths
1, i.e.\ all contributions from clusters, whose linear extension perpendicular
to the ground state planes is 1. $f_{0}$ denotes the sum of all cluster
contributions of widths 2, which are not cut by a ground state plane, 
$f_{1}$ denotes the sum of all cluster contributions of widths 2, which are
cut by a ground state plane and so on. A systematic low temperature expansion
of the cluster functions $f_{q}$ can now be generated by analysing the
corresponding cluster contributions. We obtain
\begin{eqnarray}
c^{(g)}(\I) & = & \sum_{\J\subseteq\I} (-1)^{(|\I|-|\J|)}\;\ln\Big(
\sum_{\K\subseteq\J}z^{(g)}(\K)\Big) \\[3mm]
& = & \sum_{r=1}^{\infty} \frac{(-1)^{(r-1)}}{r} \sum_{\parbox[t]{1.9cm}{$
\scriptstyle \K_{1}\cup\ldots\cup\K_{r}=\I \\
\K_{i}\subseteq\I, \:\K_{i}\not=\emptyset$}} \:\:
\prod_{j=1}^{r} z^{(g)}(\K_{j}) \nonumber \\[3mm] 
\label{entwicklung}
& = & z^{(g)}(\I)-\frac{1}{2}\; \sum_{\parbox[t]{1.4cm}{$ \scriptstyle 
\K_{1}\cup\K_{2}=\I \\ \K_{1},\K_{2}\subseteq\I \\ 
\K_{1},\K_{2}\not=\emptyset$}} \:\:
z^{(g)}(\K_{1})z^{(g)}(\K_{2}) \quad + \quad \ldots
\end{eqnarray}
The Boltzmann factors $z^{(g)}(\I)$ are of the form:
\begin{eqnarray}
z^{(g)}(\I) & = & x^{n_{2}}y^{n_{4}}, \\[1mm] \nonumber
\mbox{where} && \begin{array}[t]{ll}
x=\mbox{e}^{-\beta J}, & n_{2}=\mbox{Number of simple edges} \\
y=\mbox{e}^{-\beta 4kJ}, & n_{4}=\mbox{Number of double edges.} 
\end{array}
\end{eqnarray}
From here one can see, that a cluster $\I$ contributes to an order in
$x$, which is higher or equal to $4(b_{1}(\I)+b_{2}(\I)+b_{3}(\I))$, if
$b_{i}(\I)$ denote the side lengths of the smallest box, which contains $\I$.
Thus to generate an expansion in $x$ up to order $n$, only clusters $\I$
with $4(b_{1}(\I)+b_{2}(\I)+b_{3}(\I))\leq n$ have to be considered. \par
\begin{figure}[htb]
\begin{center}
\begin{tabular}{cc}
\includegraphics[angle=270,width=6.5cm]{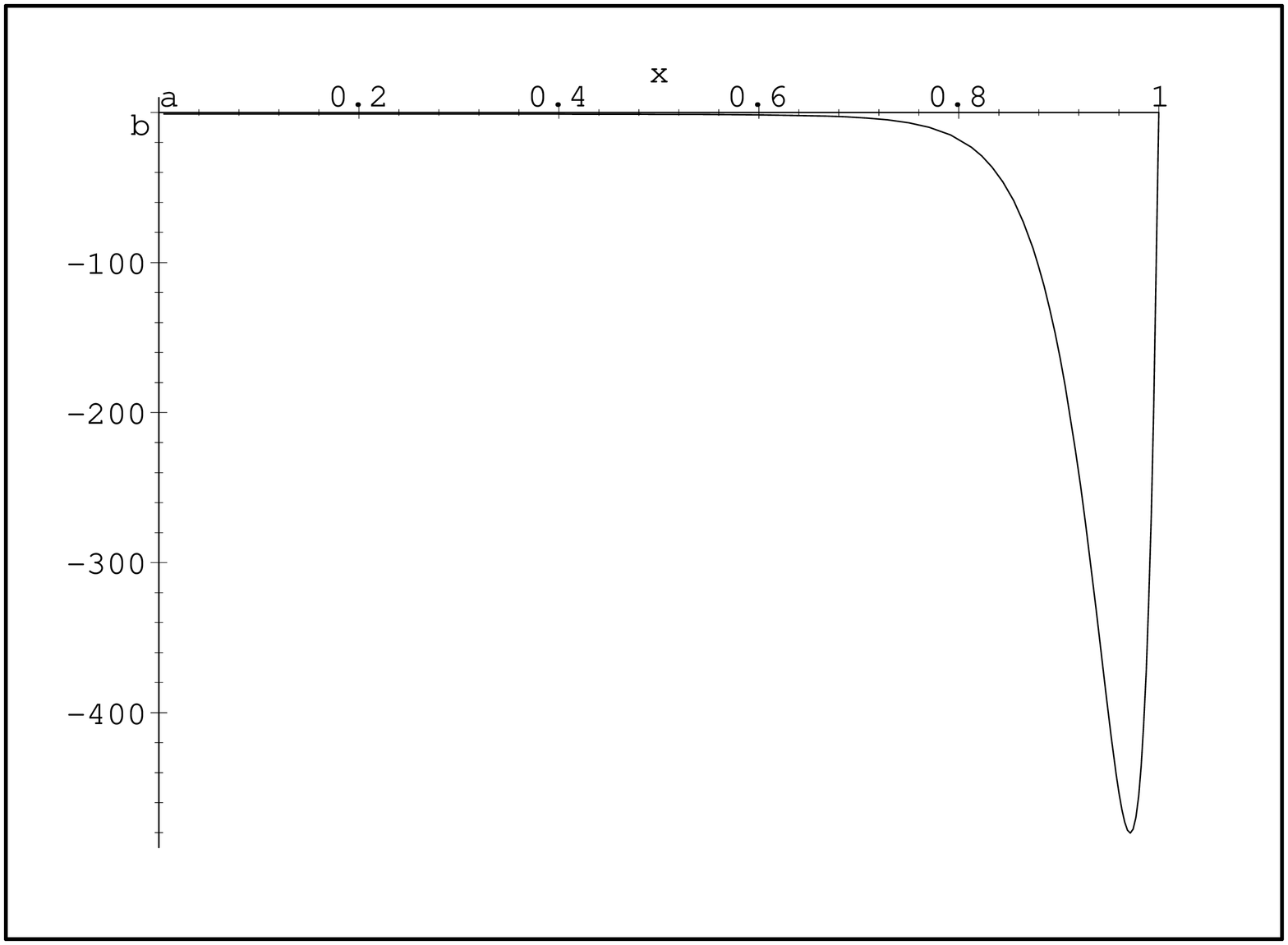} 
\includegraphics[angle=270,width=6.5cm]{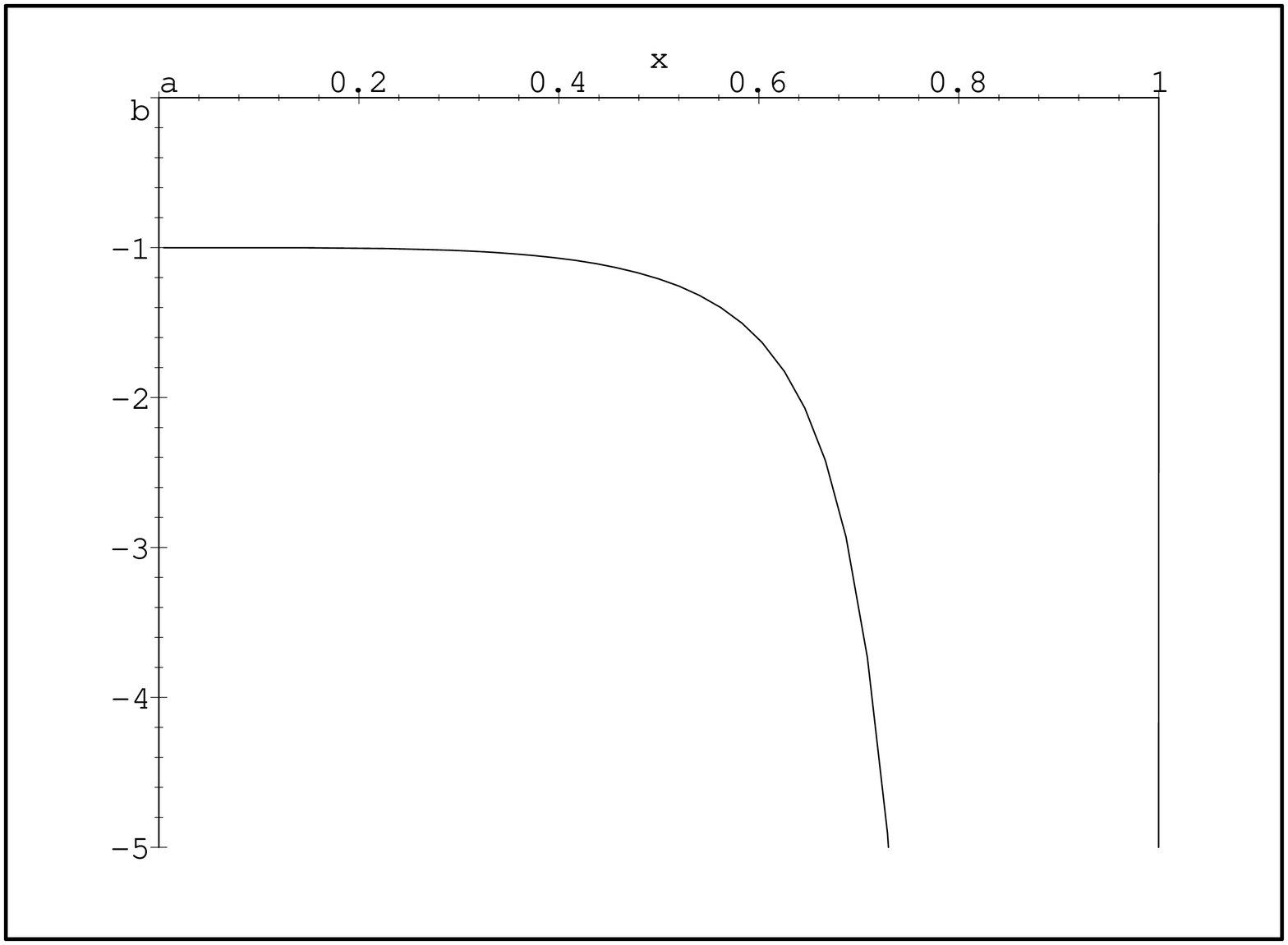} 
\end{tabular}
\end{center}
\caption{\label{f1} $(f_{1}-f_{0})/x^{16}$ is plotted as a function of
$x=\mbox{e}^{-\beta J}$ for $k=1$.}
\end{figure}

With the method described above, we calculated the cluster functions
$f_{q}$ up to order $x^{38}$. The numbers of clusters which contribute to a 
specific order are listed in table \ref{anzahl}.
\begin{table}[ht]
\begin{center}
{\renewcommand{\arraystretch}{1.3}
\begin{tabular}{|c|c|} \hline
Number of simple edges & Number of clusters \\ \hline \hline
12 & 1      \\ \hline
14 & 0      \\ \hline
16 & 3      \\ \hline
18 & 0      \\ \hline
20 & 6      \\ \hline
22 & 18     \\ \hline
24 & 10     \\ \hline
26 & 96     \\ \hline
28 & 105    \\ \hline
30 & 372    \\ \hline
32 & 789    \\ \hline
34 & 1806   \\ \hline
36 & 4881   \\ \hline
38 & 10134  \\ \hline
\end{tabular}}
\caption{\label{anzahl} Numbers of connected spin clusters with a given
number of simple edges.}
\end{center}
\end{table} 
To see, whether or not layered low temperature phases arise, we write according
to (\ref{zerlegung}):
\begin{equation}
\label{stabilitaet}
\frac{Z_{N}}{2Z_{N}^{(0)}}=1+3\sum_{g=1}^{2^{L-1}-1}
\frac{Z_{N}^{(g)}}{Z_{N}^{(0)}}.
\end{equation} 
Using (\ref{darstellung}) we arrive at:
\begin{eqnarray}
\nonumber
\lim\limits_{L\to\infty} \frac{Z_{N}}{2Z_{N}^{(0)}} & = & 
1+3 \lim\limits_{L\to\infty} \sum\limits_{g=1}^{2^{L-1}-1}
\mbox{exp}\Bigg(L^{2} \Big(a_{1}(g)(f_{1}-f_{0})+
a_{01}(g)(f_{01}-f_{00}) \\[1mm] 
\label{stabilitaet1}
&& +a_{10}(g)(f_{10}-f_{00})+a_{11}(g)(f_{11}-f_{00})+\ldots\Big) \Bigg).
\end{eqnarray}
Since the functions $(f_{1}-f_{0})$,$(f_{01}-f_{00})$,$\ldots$ turn out to
be negative in the range $0<x<1$ and vanish only at $x=0$ and $x=1$ 
for all positive $k$, the right hand side of (\ref{stabilitaet1}) tends to 1 
in the thermodynamical limit. It follows that layered phases are 
suppressed at low but non-zero temperature. In figure
\ref{f1} the function $(f_{1}-f_{0})/x^{16}$ is plotted for $k=1$.
Higher order functions $(f_{01}-f_{00})/x^{20}$,$\ldots$ show the same
qualitative behaviour.
The total free energy $F$ can therefore be calculated
as a low temperature expansion, where only excitations of the ferromagnetic
ground states are considered, i.e.\ 
\begin{equation}
\label{f_energie}
-\beta F =\mbox{ln}(2) + \mbox{ln}(Z^{(0)}) = \mbox{ln}(2) + f + f_{0} 
+ f_{00} + f_{000} + \ldots
\end{equation}
To order $x^{38}$ this yields:
\begin{eqnarray*}
-\beta F & = & \mbox{ln}(2)+
x^{12}+3x^{16}+6x^{20}+(6y+12)x^{22}+\frac{1}{2}x^{24}+(48+48y)x^{26} \\&&
+(15+6y^{2})x^{28}+(136+204y+24y^{2}+8y^{3})x^{30}+(228y+90y^{2} \\&&
+\frac{129}{2})x^{32}+(456y+216y^{2}+78y^{3}+456)x^{34}+(1824y+\frac{1975}{3} \\
&& +1032y^{2}+56y^{3}+24y^{4}+2y^{6})x^{36}+(1110y+1224y^{2}+600y^{3} \\&&
+108y^{4}+540+24y^{5})x^{38}+\ldots
\end{eqnarray*}
where $x=\mbox{e}^{-\beta J}$ and $y=\mbox{e}^{-\beta 4kJ}$. With the
formalism described above, it is also possible to derive a low temperature
expansion of the surface tension $\sigma$, which is defined as:
\begin{eqnarray}
\label{sigmadef}
\sigma & := & \lim_{L\to\infty} \frac{1}{L^{2}}\:\: \mbox{ln}\left(
\frac{Z_{N}^{(0)}}{Z_{N}^{(1)}} \right) \\[3mm]
& = & (f_{0}-f_{1})+(f_{00}-f_{01})
+(f_{00}-f_{10})+(f_{000}-f_{001})+\dots
\end{eqnarray}
In this expression $Z_{N}^{(1)}$ denotes the contribution of the total 
partition function from a ground state which contains exactly one plane domain
wall. Clearly $\sigma$ vanishes for $T=0$. However for finite $T$ a finite 
surface tension is generated by thermal fluctuation. We find the following 
series expansion: 
\begin{eqnarray}
\label{sigma}
\sigma & = &
(1-y^{4})x^{16}+(-2y^{4}-2y^{6}+4)x^{20}+(4y+4-8y^{3})x^{22}+(-4y^{6} \\&&
\nonumber
+10-3y^{4}-3y^{8})x^{24}+(16+48y-8y^{6}-24y^{5}-24y^{3}-8y^{4})x^{26} \\&&
\nonumber 
+(-10y^{8}+72-32y^{3}-16y^{5}-6y^{10}-18y^{2}+16y^{4}-6y^{6})x^{28} \\&&
\nonumber
+(32+264y+16y^{2}-40y^{3}-48y^{4}-40y^{6}-120y^{5}-16y^{8}-48y^{7})x^{30} \\&&
\nonumber
+(240y+4y^{2}-200y^{3}-27y^{4}-4y^{6}-96y^{5}-\frac{65}{2}y^{8}-32y^{10}-40y^{7}
\\&& \nonumber
-13y^{12}+\frac{401}{2})x^{32}+(828y+104y^{2}+48y^{3}-304y^{4}-160y^{6}-540y^{5}
\\&& \nonumber
-136y^{8}-24y^{10}-296y^{7}-96y^{9}-4y^{11}+580)x^{34}+(2208y+1114y^{2} \\&&
\nonumber
-792y^{3}-718y^{4}-344y^{6}-928y^{5}-146y^{8}-88y^{10}-424y^{7}-90y^{12} \\&&
\nonumber
-64y^{9}-16y^{11}-26y^{14}+314)x^{36}+(3172y+568y^{2}+632y^{3}-1056y^{4} \\&&
\nonumber
-1272y^{6}-1268y^{5}-712y^{8}-352y^{10}-1296y^{7}-64y^{12}-748y^{9} \\&&
\nonumber
-232y^{11}-16y^{13}+2644)x^{38}+\ldots 
\end{eqnarray}
\section{Critical exponents}
If the self-avoiding coupling $k$ is sufficiently large ($k>0.3$), the authors of
\cite{johnston4,johnston5} find a second order phase transition, whereas at 
$k=0$ the transition is of first order. These results were obtained by
simulations. Using the result above, i.e. that only the ferromagnetically
ordered phases contribute to the free energy, we were able to derive
low temperature expansions for the magnetization, susceptibility and specific 
heat up to order $x^{44}$ for fixed values of $k$ 
($k=\frac{n}{4}, n=0,\ldots,12$). A Pad\'e analysis yields critical exponents,
which are in agreement with previous results. \par
In the following we describe the method used to derive the series expansions
\cite{enting1,enting2}. The free energy can be derived from the ferromagnetic 
part $Z^{(0)}$ of the partition function. According to (\ref{clustersumme}) we 
write:
\begin{equation}
\mbox{ln}(Z^{(0)}_{N}) = \sum_{\I\subseteq\N}c^{(0)}(\I).
\end{equation}
To each cluster $\I$ of overturned spins corresponds a minimal box $q$ with
side lengths $q_{1},q_{2},q_{3}$, which contains $\I$. We define
\begin{equation}
C(q):=\parbox[t]{10cm}{Sum of all cluster contributions $c^{(0)}(\I)$ modulo
translations, such that $q$ is the minimal box of $\I$.}
\end{equation} 
If $Z(Q)$ denotes the partition function for the finite box $Q$, evaluated
for a ferromagnetic ground state, the following equation holds due to
translation invariance:
\begin{equation}
\label{inversion}
\mbox{ln}(Z(Q))=\sum_{q\subseteq Q} (Q_{1}-q_{1}+1)(Q_{2}-q_{2}+1)
(Q_{3}-q_{3}+1)C(q).
\end{equation}
Therefore the free energy can be written as:
\begin{equation}
\label{trunkierung}
-\beta F =\sum_{q}C(q)=
\sum_{q\,:\: 4(q_{1}+q_{2}+q_{3})<=n}C(q) + O(x^{n+2}).
\end{equation}
Inverting equation (\ref{inversion}) and putting the resulting expression for 
$C(q)$ in (\ref{trunkierung}), we finally arrive at:
\begin{eqnarray}
\label{trunkierung1}
-\beta F & = & \sum_{Q\,:\: Q_{1}+Q_{2}+Q_{3}<=l } 
\left(\scriptstyle 5\atop \scriptstyle l-Q_{1}-Q_{2}-Q_{3} \right)\,
(-1)^{l-Q_{1}-Q_{2}-Q_{3}} \;\mbox{ln}(Z(Q)) \\[2mm]
&& + O(x^{4l+2}), \nonumber
\end{eqnarray} 
where $\left(\scriptstyle 5\atop \scriptstyle n\right)=0$,
if $n<0$ or $n>5$.
Thus in order to generate a low temperature expansion for $F$, we just need to
calculate partition functions for small volumes, which can be done very
efficiently by recursive methods\footnote{A detailed description of the used
algorithm can be found in \cite{diss}.} \cite{binder1}. Expanding
\begin{eqnarray}
\label{konfigsumme}
Z(Q) & = & \sum_{\I\subseteq\N}x^{n_{2}(\I)+4k\,n_{4}(\I)}\,z^{n(\I)} \\[2mm]
& = & \Lambda^{(Q)}_{0}(x)+\Lambda^{(Q)}_{1}(x)(z-1) 
+\Lambda^{(Q)}_{2}(x) (z-1)^{2} +\ldots,
\end{eqnarray}
where $n(\I)$ denotes the number of spins in $\I$, series expansions for the 
magnetization $M(x)$, susceptibility $S(x)$ and specific heat $C(x)$ 
can be derived from the polynomials $\Lambda^{(Q)}_{0}(x)$,
$\Lambda^{(Q)}_{1}(x)$,$\Lambda^{(Q)}_{2}(x)$ in the following way:
\begin{eqnarray}
M(x) & = & 1-2\left.\frac{\partial(-\beta F)}{\partial z} \right|_{z=1} \\[2mm]
S(x) & = & \left.z\frac{\partial}{\partial z}\Big(z
\frac{\partial(-\beta F)}{\partial z}\Big)\right|_{z=1} \\[2mm]
C(x) & = & \left.x\frac{\partial}{\partial x}\Big(x
\frac{\partial(-\beta F)}{\partial x}\Big)\right|_{z=1}.
\end{eqnarray}  
The resulting expansions for fixed values of $k=\frac{n}{4}, n=0,\ldots ,12$ 
can be found in appendix A. To calculate critical exponents, we analysed 
$[n/(n-2)],[n/(n-1)],[n/n],[n/(n+1)]$ and $[n/(n+2)]$ unbiased Dlog Pad\'e
approximants of the corresponding series 
expansions \cite{guttmann}. The results are shown in table 
\ref{beta}, \ref{gamma} and \ref{alpha}. The errors roughly estimate the
fluctuations within the Pad\'e tables. Defective approximants have been 
excluded from the analysis.
\begin{table}[htb]
\begin{center}
\begin{tabular}{|>{$}c<{$}||>{$}c<{$}|>{$}c<{$}|} \hline
k & x_{c} & \beta \\ \hline \hline
0.50 & 0.636\pm 0.002 & 0.037\pm 0.002 \\ \hline
0.75 & 0.642\pm 0.003 & 0.045\pm 0.004 \\ \hline
1.00 & 0.641\pm 0.001 & 0.040\pm 0.002 \\ \hline
1.25 & 0.644\pm 0.001 & 0.045\pm 0.002 \\ \hline
1.50 & 0.646\pm 0.001 & 0.047\pm 0.002 \\ \hline
1.75 & 0.647\pm 0.002 & 0.049\pm 0.003 \\ \hline
2.00 & 0.647\pm 0.002 & 0.050\pm 0.004 \\ \hline
2.25 & 0.647\pm 0.002 & 0.048\pm 0.007 \\ \hline
2.50 & 0.648\pm 0.003 & 0.050\pm 0.006 \\ \hline
2.75 & 0.648\pm 0.002 & 0.051\pm 0.006 \\ \hline
3.00 & 0.649\pm 0.002 & 0.052\pm 0.004 \\ \hline
\end{tabular}
\caption{\label{beta} Critical temperature $x_{c}=\mbox{e}^{-\beta_{c}J}$
and magnetization exponent $\beta$.} 
\end{center}
\end{table}
In \cite{johnston4} the critical $x$ was determined as $x_{c}\approx 0.64$.
A cluster variation-Pad\'e calculation also confirms this result 
\cite{pelizzola1}. The authors determine the critical exponent of the
magnetization as $\beta=0.062\pm 0.003$ at a critical temperature 
$x_{c}\approx 0.65$. 
\begin{table}[htb]
\begin{center}
\begin{tabular}{|>{$}c<{$}||>{$}c<{$}|>{$}c<{$}|} \hline
k & x_{c} & \gamma \\ \hline \hline
0.50 & 0.642\pm 0.002  & 1.7\pm 0.1 \\ \hline
0.75 & 0.647\pm 0.008  & 1.8\pm 0.3   \\ \hline
1.00 & 0.64\pm 0.01    & 1.7\pm 0.2   \\ \hline
1.25 & 0.65\pm 0.01    & 1.7\pm 0.3   \\ \hline
1.50 & 0.65\pm 0.01    & 1.7\pm 0.3   \\ \hline
1.75 & 0.65\pm 0.01    & 1.6\pm 0.3   \\ \hline
2.00 & 0.65\pm 0.01    & 1.7\pm 0.3   \\ \hline
2.25 & 0.64\pm 0.02    & 1.7\pm 0.3   \\ \hline
2.50 & 0.65\pm 0.02    & 1.7\pm 0.7   \\ \hline
2.75 & 0.64\pm 0.02    & 1.5\pm 0.8   \\ \hline
3.00 & 0.65\pm 0.02    & 1.3\pm 0.6   \\ \hline
\end{tabular}
\caption{\label{gamma} Critical temperature $x_{c}=\mbox{e}^{-\beta_{c}J}$
and susceptibility exponent $\gamma$.} 
\end{center}
\end{table}
The exponent $\gamma$ was also calculated in \cite{johnston4} as 
$\gamma\approx 1.6$. An estimate of $\gamma\approx 1.4$ was given in 
\cite{pelizzola2}. Our values are of the same order of magnitude. However,
the fluctuations within the Pad\'e tables are relatively large. Unfortunately
different extrapolation techniques like inhomogeneous Pad\'e approximants
or differential approximants did not yield better results. 
\begin{table}[htb]
\begin{center}
\begin{tabular}{|>{$}c<{$}||>{$}c<{$}|>{$}c<{$}|} \hline
k & x_{c} & \alpha \\ \hline \hline
0.50 & 0.647\pm 0.002 & 0.83\pm 0.03 \\ \hline
0.75 & 0.646\pm 0.004 & 0.67\pm 0.08 \\ \hline
1.00 & 0.65\pm  0.01  & 0.62\pm 0.03 \\ \hline
1.25 & 0.649\pm 0.001 & 0.57\pm 0.01 \\ \hline
1.50 & 0.649\pm 0.001 & 0.53\pm 0.01 \\ \hline
1.75 & 0.648\pm 0.002 & 0.49\pm 0.02 \\ \hline
2.00 & 0.648\pm 0.001 & 0.45\pm 0.02 \\ \hline
2.25 & 0.64\pm 0.03   & 0.5\pm 0.1   \\ \hline
2.50 & 0.647\pm 0.003 & 0.41\pm 0.06 \\ \hline
2.75 & 0.64\pm 0.03   & 0.4\pm 0.2   \\ \hline
3.00 & 0.63\pm 0.03   & 0.3\pm 0.2   \\ \hline
\end{tabular} 
\caption{\label{alpha} Critical temperature $x_{c}=\mbox{e}^{-\beta_{c}J}$
and specific heat exponent $\alpha$.}  
\end{center}
\end{table}
The exponent $\alpha$ of the specific heat was determined in \cite{johnston6}
as $\alpha\approx 0.7$. Our values agree roughly with this result. Since
the errors are quite large also in this case, it is unclear whether the
exponents are $k$-dependent or not. Furthermore we analysed the series expansion
of the surface tension $\sigma(x,y)$ as given in (\ref{sigma}). The surface 
tension should scale at the critical point and vanish for higher temperatures.
Surprisingly, the Pad\'e analysis of the series expansion gives negative
values for the corresponding exponent $\mu$. Moreover, $\sigma(x,y)$ can be
written as:
\begin{equation}
\sigma(x,y)=(1-y)P(x,y),
\end{equation}
where all coefficients in the expansion of $P(x,y)$ are positive. This 
indicates a diverging surface tension, which is of course not the expected
behaviour. However a similar situation arises in the ordinary 3-dimensional
Ising model. The low temperature expansion of the surface tension \cite{shaw}
reads in this case:
\begin{eqnarray*}
\sigma(x) & = & -\mbox{ln}(x)-2\,x^{4}-2\,x^{6}-10\,x^{8}-16\,x^{10}-
\frac{242}{3}\,x^{12}-150\,x^{14} \\
&& -734\,x^{16}-\frac{4334}{3}\,x^{18}-\ldots \qquad (x=\mbox{e}^{-2\beta J}). 
\end{eqnarray*}
Here the expansion coefficients all have the same sign and a Pad\'e analysis
also yields negative critical exponents. Nevertheless, simulations indicate,
that the surface tension scales at the critical point \cite{pinn}. 
The corresponding exponent is positive and consistent with Widom-scaling. The
reason for this strange behaviour lies in the existence of a roughening
transition at a temperature $T_{R}$ below $T_{c}$ \cite{pinn}. 
Renormalization group calculations indicate, that this transition is of 
Kosterlitz-Thouless type \cite{hasenbusch} with an essential singularity  
at $T_{R}$. We take this as a strong hint, that a similar transition occurs
in the surface tension of the gonihedric Ising model. This point needs however
further clarification.
\section{Summary}
The gonihedric Ising model was introduced as a lattice discretization of
the so called gonihedric string. The gonihedric string itself is a model for 
triangulated random surfaces. Whether or not both theories are equivalent in the
sense that they are different discretizations of the same continuum random 
surface theory is not obvious. A necessary condition for the existence of a 
continuum limit is the occurrence of a second order phase transition. The
existence of ordered low temperature phases often signals such a phase 
transition, if at high temperatures a disordered phase is expected. \par
In this paper we studied some low temperature properties of the gonihedric 
Ising model. If $k=0$, the model shows apart from the global 
$\Z_{2}$ symmetry an additional layer-flip symmetry. With a suitable 
extension of Peierls contour method we proved, that the 
layer-flip symmetry is spontaneously broken at low temperatures for $d\geq 3$. 
We found infinitely many ordered low temperature phases, one for each ground 
state. However, simulations indicate, that the corresponding phase transition
is of first order \cite{johnston5}. Thus in this case a continuum limit does
not exist. For $k>0.3$ the authors of 
\cite{savvidy10,johnston4,johnston6,pelizzola1}
find a second order phase transition. It was suggested in \cite{pelizzola1},
that in contrast to the case $k=0$ only the ferromagnetic phases remain stable,
while layered phases are thermodynamically suppressed. The low temperature
expansion of the free energy in section 2 supports this picture. For $k>0$, a
finite surface tension is generated if $T>0$, i.e. the occurrence of layers
is energetically suppressed by a factor $\mbox{e}^{-\beta L^2}$. On the other
hand the ground state entropy grows only like $2^{L}$ with the linear size $L$ 
of the system and can therefore not compensate the energy factor. Using this 
result, we calculated low temperature expansions of the magnetization, 
susceptibility and specific heat, where only excitations of the ferromagnetic 
ground states were considered. The critical exponents, derived from Pad\'e 
approximants are in agreement with previous results. Unfortunately the 
fluctuations within the Pad\'e tables are quite large (in particular for 
$\gamma$ and $\alpha$). It is therefore not possible to determine, whether or 
not the model shows non-universal behaviour. Finally we analysed the surface 
tension. A Pad\'e-approximation resulted in a negative critical
exponent. This probably indicates the existence of a roughening transition
as in the ordinary 3-dimensional Ising model.
\begin{appendix}
\section{Series expansions}
This appendix contains the coefficients of the series expansions, used in
section 4 to calculate critical exponents. The corresponding Pad\'e tables
can be found in \cite{diss}.
\setlength{\tabcolsep}{1mm}
\renewcommand{\arraystretch}{1.1}
\begin{small}
\subsection{Magnetization}
\begin{center}
\begin{longtable}{|c|c|c|c|c|c|c|c|} \hline
$k$ & 0 & 0.25 & 0.5 & 0.75 & 1 & 1.25 & 1.5 \\ \hline \hline
$x^{0}$ & 1 & 1 & 1 & 1 & 1 & 1 & 1 \\ \hline
$x^{12}$ & -2 & -2 & -2 & -2 & -2 & -2 & -2  \\ \hline 
$x^{16}$ & -12 & -12 & -12 & -12 & -12 & -12 & -12  \\ \hline 
$x^{20}$ & -42 & -42 & -42 & -42 & -42 & -42 & -42  \\ \hline 
$x^{22}$ & -96 & -72 & -72 & -72 & -72 & -72 & -72  \\ \hline 
$x^{23}$ & 0 & -24 & 0 & 0 & 0 & 0 & 0  \\ \hline 
$x^{24}$ & -74 & -74 & -98 & -74 & -74 & -74 & -74  \\ \hline 
$x^{25}$ & 0 & 0 & 0 & -24 & 0 & 0 & 0  \\ \hline 
$x^{26}$ & -720 & -432 & -432 & -432 & -456 & -432 & -432  \\ \hline 
$x^{27}$ & 0 & -288 & 0 & 0 & 0 & -24 & 0  \\ \hline 
$x^{28}$ & -564 & -516 & -804 & -516 & -516 & -516 & -540  \\ \hline 
$x^{29}$ & 0 & 0 & 0 & -288 & 0 & 0 & 0  \\ \hline 
$x^{30}$ & -3680 & -1736 & -1688 & -1688 & -1976 & -1688 & -1688  \\ \hline 
$x^{31}$ & 0 & -1752 & 0 & 0 & 0 & -288 & 0  \\ \hline 
$x^{32}$ & -5316 & -2856 & -4464 & -2664 & -2664 & -2664 & -2952  \\ \hline 
$x^{33}$ & 0 & -1920 & 0 & -1752 & 0 & 0 & 0  \\ \hline 
$x^{34}$ & -18048 & -9468 & -10752 & -8736 & -10440 & -8688 & -8688  \\ \hline 
$x^{35}$ & 0 & -6456 & 0 & -1872 & 0 & -1752 & 0  \\ \hline 
$x^{36}$ & -46046 & -17582 & -22610 & -15518 & -17246 & -15326 & -17078  \\ \hline 
$x^{37}$ & 0 & -20184 & 0 & -6456 & 0 & -1872 & 0  \\ \hline 
$x^{38}$ & -84696 & -41472 & -52992 & -31980 & -37848 & -31248 & -33072  \\ \hline 
$x^{39}$ & 0 & -29840 & 0 & -19584 & 0 & -6456 & 0  \\ \hline 
$x^{40}$ & -337326 & -118098 & -141546 & -103698 & -121758 & -101634 & -107946  \\ \hline 
$x^{41}$ & 0 & -128760 & 0 & -29184 & 0 & -19536 & 0  \\ \hline 
$x^{42}$ & -518832 & -236876 & -281280 & -152344 & -173560 & -142852 & -161800  \\ \hline 
$x^{43}$ & 0 & -174816 & 0 & -122784 & 0 & -29184 & 0  \\ \hline 
$x^{44}$ & -2016750 & -619662 & -739530 & -496470 & -612462 & -482310 & -510018  \\ \hline 
$x^{45}$ & 0 & -731672 & 0 & -159560 & 0 & -122184 & 0  \\ \hline 
\end{longtable} 
\vspace{3mm}
\begin{longtable}{|c|c|c|c|c|c|c|} \hline
$k$ & 1.75 & 2 & 2.25 & 2.5 & 2.75 & 3  \\ \hline \hline
$x^{0}$ & 1 & 1 & 1 & 1 & 1 & 1   \\ \hline
$x^{12}$ & -2 & -2 & -2 & -2 & -2 & -2  \\ \hline 
$x^{16}$ & -12 & -12 & -12 & -12 & -12 & -12  \\ \hline 
$x^{20}$ & -42 & -42 & -42 & -42 & -42 & -42  \\ \hline 
$x^{22}$ & -72 & -72 & -72 & -72 & -72 & -72  \\ \hline 
$x^{24}$ & -74 & -74 & -74 & -74 & -74 & -74  \\ \hline 
$x^{26}$ & -432 & -432 & -432 & -432 & -432 & -432  \\ \hline 
$x^{28}$ & -516 & -516 & -516 & -516 & -516 & -516  \\ \hline 
$x^{29}$ & -24 & 0 & 0 & 0 & 0 & 0  \\ \hline 
$x^{30}$ & -1688 & -1712 & -1688 & -1688 & -1688 & -1688  \\ \hline 
$x^{31}$ & 0 & 0 & -24 & 0 & 0 & 0  \\ \hline 
$x^{32}$ & -2664 & -2664 & -2664 & -2688 & -2664 & -2664  \\ \hline 
$x^{33}$ & -288 & 0 & 0 & 0 & -24 & 0  \\ \hline 
$x^{34}$ & -8688 & -8976 & -8688 & -8688 & -8688 & -8712  \\ \hline 
$x^{35}$ & 0 & 0 & -288 & 0 & 0 & 0  \\ \hline 
$x^{36}$ & -15326 & -15326 & -15326 & -15614 & -15326 & -15326  \\ \hline 
$x^{37}$ & -1752 & 0 & 0 & 0 & -288 & 0  \\ \hline 
$x^{38}$ & -31200 & -32952 & -31200 & -31200 & -31200 & -31488  \\ \hline 
$x^{39}$ & -1872 & 0 & -1752 & 0 & 0 & 0  \\ \hline 
$x^{40}$ & -101442 & -103314 & -101442 & -103194 & -101442 & -101442  \\ \hline 
$x^{41}$ & -6456 & 0 & -1872 & 0 & -1752 & 0  \\ \hline 
$x^{42}$ & -142120 & -148528 & -142072 & -143944 & -142072 & -143824  \\ \hline 
$x^{43}$ & -19536 & 0 & -6456 & 0 & -1872 & 0  \\ \hline 
$x^{44}$ & -480246 & -499638 & -480054 & -486510 & -480054 & -481926  \\ \hline 
$x^{45}$ & -29184 & 0 & -19536 & 0 & -6456 & 0  \\ \hline 
\end{longtable}
\end{center}
\subsection{Susceptibility}
\begin{center}
\begin{longtable}{|c|c|c|c|c|c|c|c|} \hline
$k$ & 0 & 0.25 & 0.5 & 0.75 & 1 & 1.25 & 1.5 \\ \hline \hline
$x^{12}$ & 1 & 1 & 1 & 1 & 1 & 1 & 1  \\ \hline  
$x^{16}$ & 12 & 12 & 12 & 12 & 12 & 12 & 12  \\ \hline 
$x^{20}$ & 75 & 75 & 75 & 75 & 75 & 75 & 75  \\ \hline 
$x^{22}$ & 132 & 108 & 108 & 108 & 108 & 108 & 108  \\ \hline 
$x^{23}$ & 0 & 24 & 0 & 0 & 0 & 0 & 0  \\ \hline 
$x^{24}$ & 290 & 290 & 314 & 290 & 290 & 290 & 290  \\ \hline 
$x^{25}$ & 0 & 0 & 0 & 24 & 0 & 0 & 0  \\ \hline 
$x^{26}$ & 1416 & 984 & 984 & 984 & 1008 & 984 & 984  \\ \hline 
$x^{27}$ & 0 & 432 & 0 & 0 & 0 & 24 & 0  \\ \hline 
$x^{28}$ & 2274 & 2178 & 2610 & 2178 & 2178 & 2178 & 2202  \\ \hline 
$x^{29}$ & 0 & 0 & 0 & 432 & 0 & 0 & 0  \\ \hline 
$x^{30}$ & 9664 & 5500 & 5404 & 5404 & 5836 & 5404 & 5404  \\ \hline 
$x^{31}$ & 0 & 3804 & 0 & 0 & 0 & 432 & 0  \\ \hline 
$x^{32}$ & 20064 & 14742 & 18258 & 14358 & 14358 & 14358 & 14790  \\ \hline 
$x^{33}$ & 0 & 3936 & 0 & 3804 & 0 & 0 & 0  \\ \hline 
$x^{34}$ & 65916 & 39138 & 41544 & 37392 & 41100 & 37296 & 37296  \\ \hline 
$x^{35}$ & 0 & 21324 & 0 & 3864 & 0 & 3804 & 0  \\ \hline 
$x^{36}$ & 170779 & 93603 & 110913 & 88059 & 91635 & 87675 & 91479  \\ \hline 
$x^{37}$ & 0 & 54192 & 0 & 21324 & 0 & 3864 & 0  \\ \hline 
$x^{38}$ & 420492 & 233664 & 264696 & 207786 & 227652 & 206040 & 209808  \\ \hline 
$x^{39}$ & 0 & 138896 & 0 & 52896 & 0 & 21324 & 0  \\ \hline 
$x^{40}$ & 1400784 & 660426 & 770154 & 610026 & 658764 & 604482 & 625518  \\ \hline 
$x^{41}$ & 0 & 445740 & 0 & 136968 & 0 & 52824 & 0  \\ \hline 
$x^{42}$ & 2963424 & 1526998 & 1701136 & 1244228 & 1359476 & 1218350 & 1269716  \\ \hline 
$x^{43}$ & 0 & 983064 & 0 & 428340 & 0 & 136968 & 0  \\ \hline 
$x^{44}$ & 9979599 & 3989439 & 4711941 & 3507195 & 3906159 & 3457395 & 3590277  \\ \hline 
$x^{45}$ & 0 & 3297892 & 0 & 935276 & 0 & 427044 & 0  \\ \hline 
\end{longtable} 
\vspace{3mm}
\begin{longtable}{|c|c|c|c|c|c|c|} \hline
$k$ & 1.75 & 2 & 2.25 & 2.5 & 2.75 & 3  \\ \hline \hline
$x^{12}$ & 1 & 1 & 1 & 1 & 1 & 1  \\ \hline 
$x^{16}$ & 12 & 12 & 12 & 12 & 12 & 12  \\ \hline 
$x^{20}$ & 75 & 75 & 75 & 75 & 75 & 75  \\ \hline 
$x^{22}$ & 108 & 108 & 108 & 108 & 108 & 108  \\ \hline 
$x^{24}$ & 290 & 290 & 290 & 290 & 290 & 290  \\ \hline 
$x^{26}$ & 984 & 984 & 984 & 984 & 984 & 984  \\ \hline 
$x^{28}$ & 2178 & 2178 & 2178 & 2178 & 2178 & 2178  \\ \hline 
$x^{29}$ & 24 & 0 & 0 & 0 & 0 & 0  \\ \hline 
$x^{30}$ & 5404 & 5428 & 5404 & 5404 & 5404 & 5404  \\ \hline 
$x^{31}$ & 0 & 0 & 24 & 0 & 0 & 0  \\ \hline 
$x^{32}$ & 14358 & 14358 & 14358 & 14382 & 14358 & 14358  \\ \hline 
$x^{33}$ & 432 & 0 & 0 & 0 & 24 & 0  \\ \hline 
$x^{34}$ & 37296 & 37728 & 37296 & 37296 & 37296 & 37320  \\ \hline 
$x^{35}$ & 0 & 0 & 432 & 0 & 0 & 0  \\ \hline 
$x^{36}$ & 87675 & 87675 & 87675 & 88107 & 87675 & 87675  \\ \hline 
$x^{37}$ & 3804 & 0 & 0 & 0 & 432 & 0  \\ \hline 
$x^{38}$ & 205944 & 209748 & 205944 & 205944 & 205944 & 206376  \\ \hline 
$x^{39}$ & 3864 & 0 & 3804 & 0 & 0 & 0  \\ \hline 
$x^{40}$ & 604098 & 607962 & 604098 & 607902 & 604098 & 604098  \\ \hline 
$x^{41}$ & 21324 & 0 & 3864 & 0 & 3804 & 0  \\ \hline 
$x^{42}$ & 1216604 & 1237832 & 1216508 & 1220372 & 1216508 & 1220312  \\ \hline 
$x^{43}$ & 52824 & 0 & 21324 & 0 & 3864 & 0  \\ \hline 
$x^{44}$ & 3451851 & 3504387 & 3451467 & 3472791 & 3451467 & 3455331  \\ \hline 
$x^{45}$ & 136968 & 0 & 52824 & 0 & 21324 & 0  \\ \hline 
\end{longtable}
\end{center}
\subsection{Specific heat}
\begin{center}
\begin{longtable}{|c|c|c|c|c|c|c|c|} \hline
$k$ & 0 & 0.25 & 0.5 & 0.75 & 1 & 1.25 & 1.5 \\ \hline \hline
$x^{12}$ & 144 & 144 & 144 & 144 & 144 & 144 & 144  \\ \hline  
$x^{16}$ & 768 & 768 & 768 & 768 & 768 & 768 & 768  \\ \hline 
$x^{20}$ & 2400 & 2400 & 2400 & 2400 & 2400 & 2400 & 2400  \\ \hline 
$x^{22}$ & 8712 & 5808 & 5808 & 5808 & 5808 & 5808 & 5808  \\ \hline 
$x^{23}$ & 0 & 3174 & 0 & 0 & 0 & 0 & 0  \\ \hline 
$x^{24}$ & 288 & 288 & 3744 & 288 & 288 & 288 & 288  \\ \hline 
$x^{25}$ & 0 & 0 & 0 & 3750 & 0 & 0 & 0  \\ \hline 
$x^{26}$ & 64896 & 32448 & 32448 & 32448 & 36504 & 32448 & 32448  \\ \hline 
$x^{27}$ & 0 & 34992 & 0 & 0 & 0 & 4374 & 0  \\ \hline 
$x^{28}$ & 16464 & 11760 & 49392 & 11760 & 11760 & 11760 & 16464  \\ \hline 
$x^{29}$ & 0 & 0 & 0 & 40368 & 0 & 0 & 0  \\ \hline 
$x^{30}$ & 334800 & 127800 & 122400 & 122400 & 165600 & 122400 & 122400  \\ \hline 
$x^{31}$ & 0 & 196044 & 0 & 0 & 0 & 46128 & 0  \\ \hline 
$x^{32}$ & 391680 & 90624 & 281088 & 66048 & 66048 & 66048 & 115200  \\ \hline 
$x^{33}$ & 0 & 257004 & 0 & 222156 & 0 & 0 & 0  \\ \hline 
$x^{34}$ & 1394136 & 631176 & 818448 & 534072 & 762960 & 527136 & 527136  \\ \hline 
$x^{35}$ & 0 & 558600 & 0 & 279300 & 0 & 249900 & 0  \\ \hline 
$x^{36}$ & 4660848 & 1133136 & 1571184 & 884304 & 1156464 & 853200 & 1117584  \\ \hline 
$x^{37}$ & 0 & 2603838 & 0 & 624264 & 0 & 312132 & 0  \\ \hline 
$x^{38}$ & 5207064 & 2269968 & 3725520 & 909720 & 1472880 & 788424 & 1108992  \\ \hline 
$x^{39}$ & 0 & 1773486 & 0 & 2786472 & 0 & 693576 & 0  \\ \hline 
$x^{40}$ & 36662400 & 9576000 & 11131200 & 7924800 & 10641600 & 7617600 & 8318400  \\ \hline 
$x^{41}$ & 0 & 15956052 & 0 & 1865910 & 0 & 3066144 & 0  \\ \hline 
$x^{42}$ & 36694728 & 15734880 & 20095488 & 3972528 & 4505256 & 2310840 & 5411952  \\ \hline 
$x^{43}$ & 0 & 10472736 & 0 & 16585530 & 0 & 2052390 & 0  \\ \hline 
$x^{44}$ & 213025824 & 52614672 & 58428480 & 36509088 & 53352288 & 34557600 & 36462624  \\ \hline 
$x^{45}$ & 0 & 80842050 & 0 & 8885700 & 0 & 18022500 & 0  \\ \hline 
\end{longtable} 
\vspace{3mm}
\begin{longtable}{|c|c|c|c|c|c|c|} \hline
$k$ & 1.75 & 2 & 2.25 & 2.5 & 2.75 & 3  \\ \hline \hline
$x^{12}$ & 144 & 144 & 144 & 144 & 144 & 144  \\ \hline 
$x^{16}$ & 768 & 768 & 768 & 768 & 768 & 768  \\ \hline 
$x^{20}$ & 2400 & 2400 & 2400 & 2400 & 2400 & 2400  \\ \hline 
$x^{22}$ & 5808 & 5808 & 5808 & 5808 & 5808 & 5808  \\ \hline 
$x^{24}$ & 288 & 288 & 288 & 288 & 288 & 288  \\ \hline 
$x^{26}$ & 32448 & 32448 & 32448 & 32448 & 32448 & 32448  \\ \hline 
$x^{28}$ & 11760 & 11760 & 11760 & 11760 & 11760 & 11760  \\ \hline 
$x^{29}$ & 5046 & 0 & 0 & 0 & 0 & 0  \\ \hline 
$x^{30}$ & 122400 & 127800 & 122400 & 122400 & 122400 & 122400  \\ \hline 
$x^{31}$ & 0 & 0 & 5766 & 0 & 0 & 0  \\ \hline 
$x^{32}$ & 66048 & 66048 & 66048 & 72192 & 66048 & 66048  \\ \hline 
$x^{33}$ & 52272 & 0 & 0 & 0 & 6534 & 0  \\ \hline 
$x^{34}$ & 527136 & 582624 & 527136 & 527136 & 527136 & 534072  \\ \hline 
$x^{35}$ & 0 & 0 & 58800 & 0 & 0 & 0  \\ \hline 
$x^{36}$ & 853200 & 853200 & 853200 & 915408 & 853200 & 853200  \\ \hline 
$x^{37}$ & 279276 & 0 & 0 & 0 & 65712 & 0  \\ \hline 
$x^{38}$ & 779760 & 1074336 & 779760 & 779760 & 779760 & 849072  \\ \hline 
$x^{39}$ & 346788 & 0 & 310284 & 0 & 0 & 0  \\ \hline 
$x^{40}$ & 7579200 & 7944000 & 7579200 & 7905600 & 7579200 & 7579200  \\ \hline 
$x^{41}$ & 766536 & 0 & 383268 & 0 & 342924 & 0  \\ \hline 
$x^{42}$ & 2162664 & 2956464 & 2152080 & 2554272 & 2152080 & 2511936  \\ \hline 
$x^{43}$ & 3372576 & 0 & 843144 & 0 & 421572 & 0  \\ \hline 
$x^{44}$ & 34185888 & 37682304 & 34139424 & 35022240 & 34139424 & 34580832  \\ \hline 
$x^{45}$ & 2247750 & 0 & 3693600 & 0 & 923400 & 0  \\ \hline 
\end{longtable}
\end{center}
\end{small}
\end{appendix}

\begin{thebibliography}{99}
\bibitem{savvidy1}  R.V.\ Ambartzumian, G.K.\ Savvidy, K.G.\ Savvidy and
                    G.S.\ Sukiasian, Phys. Lett. B275 (1992) 99.
\bibitem{savvidy2}  G.K.\ Savvidy and K.G.\ Savvidy, Int. J. Mod. Phys. A8
                    (1993) 3393.
\bibitem{savvidy3}  G.K.\ Savvidy and K.G.\ Savvidy, Mod. Phys. Lett. A8 (1993)
\bibitem{savvidy4}  G.K.\ Savvidy and R.\ Schneider, Commun. Math. Phys. 16
                    (1994), 283
\bibitem{savvidy5}  G.\ Koutsoumbas, G.K.\ Savvidy and K.G.\ Savvidy, Europhys.
                    Lett. 36 (1996), 331
\bibitem{ambjorn1}  J.\ Ambj{\o}rn, B.\ Durhuus and J.\ Fr\"ohlich,
                    Nucl. Phys. B257 (1985), 433
\bibitem{durhuus}   B.\ Durhuus and T.\ Jonsson, Phys. Lett. B297 (1992), 271
\bibitem{johnston1} C.F.\ Baillie and D.A.\ Johnston, Phys. Rev. D45 (1992), 45
\bibitem{johnston2} C.F.\ Baillie, D.\ Espriu and D.A.\ Johnston, Phys. Lett.
                    B305 (1993), 109
\bibitem{johnston3} C.F.\ Baillie, A.\ Irb\"ack, W.\ Janke and D.A.\ Johnston,
                    Phys. Lett. B325 (1994), 45
\bibitem{wegner1}   G.K.\ Savvidy and F.J.\ Wegner, Nucl. Phys. B413 (1994), 605
\bibitem{wegner2}   G.K.\ Savvidy and F.J.\ Wegner, Nucl. Phys. B443 (1995), 565
\bibitem{savvidy6}  G.K.\ Savvidy and K.G.\ Savvidy, Phys. Lett. B324 (1994), 72
\bibitem{savvidy7}  G.K.\ Savvidy and K.G.\ Savvidy, Phys. Lett. B337 (1994), 
                    333
\bibitem{savvidy8}  G.K.\ Savvidy, K.G.\ Savvidy and P.G.\ Savvidy, 
                    Phys. Lett. A221 (1996), 233
\bibitem{savvidy9}  G.K.\ Savvidy and K.G.\ Savvidy, Mod. Phys. Lett. A11 
                    (1996), 1379
\bibitem{savvidy10} G.K.\ Bathas, K.G.\ Floratos, G.K.\ Savvidy and 
                    K.G.\ Savvidy, Mod. Phys. Lett. A10 (1995), 2695
\bibitem{savvidy11} G.K.\ Savvidy and K.G.\ Savvidy, Mod. Phys. Lett. A11 
                    (1996), 1379
\bibitem{karowski1} M.\ Karowski and H.J.\ Thun, Phys. Rev. Lett 54 (1985), 2556
\bibitem{karowski2} M.\ Karowski, J. Phys. A: Math. Gen. 19 (1986), 3375
\bibitem{gonnella1} E.N.M.\ Cirillo and G.\ Gonnella, J. Phys. A: Math. Gen. 28
                    (1995), 867
\bibitem{gonnella2} G.\ Gonnella, S.\ Lise and A.\ Maritan, Europhys. Lett. 32 
                    (1995), 735
\bibitem{cappi}     A.\ Cappi, P.\ Colangelo, G.\ Gonella and A.\ Maritan,
                    Nucl. Phys. B370 (1992), 659
\bibitem{baxter}    R.J.\ Baxter, {\sl Exactly solved models in statistical
                    mechanics} (Academic Press, London, 1982)
\bibitem{binder}    D.P.\ Landau and K.\ Binder, Phys. Rev. B31 (1985), 5946
\bibitem{johnston4} D.A.\ Johnston and R.P.K.C.\ Malmini, Phys. Lett. B378
                    (1996), 87
\bibitem{johnston5} D.\ Espriu, M.\ Baig, D.A.\ Johnston and R.P.K.C.\ Malmini,
                    J. Phys. A: Math. Gen. 30 (1997), 405
\bibitem{johnston6} M.\ Baig, D.\ Espriu, D.A.\ Johnston and R.P.K.C.\ Malmini,
                    {\sl String tension in gonihedric 3D Ising models}, 
                    hep-lat/9703008
\bibitem{pelizzola1}E.N.M.\ Cirillo, G.\ Gonella, D.A.\ Johnston, 
                    A.\ Pelizzola, Phys. Lett. A226 (1997), 59
\bibitem{pelizzola2}E.N.M.\ Cirillo, G.\ Gonella and A.\ Pelizzola, {\sl New
                    critical behaviour of the three-dimensional Ising model with
                    nearest-neighbour, next-nearest-neighbour and plaquette
                    interaction}, cond-mat/9612001
\bibitem{peierls}   R.\ Peierls, Proc. Cambridge Phil. Soc. 32 (1936), 477
\bibitem{griffiths} R.B.\ Griffiths, Phys. Rev. A136 (1964), 437
\bibitem{pietig}    R.\ Pietig and F.J.\ Wegner, Nucl. Phys. B466 (1996), 513
\bibitem{diss}      R.\ Pietig, Ph.D. thesis, University of Heidelberg (1997)
\bibitem{enting1}   I.G.\ Enting, Aust. J. Phys. 31 (1978), 512
\bibitem{enting2}   I.G.\ Enting, J. Phys. A: Math. Gen. 11 (1978), 563
\bibitem{binder1}   K.\ Binder, Physica 62 (1972), 508
\bibitem{guttmann}  A.J.\ Guttmann in {\sl Phase Transitions and Critical 
                    Phenomena, Vol. 13}, Academic Press Inc. (London) LTD. 
                    (1989)
\bibitem{shaw}      L.J.\ Shaw and M.E.\ Fisher, Phys. Rev. A39 (1989), 2189
\bibitem{pinn}      M.\ Hasenbusch and K.\ Pinn, Physica A192 (1993), 342
\bibitem{hasenbusch}M.\ Hasenbusch, Dissertation, Universit\"at Kaiserslautern
                    (1992)
\end{thebibliography}
\end{document}